\documentclass[useAmain sequence,usenatbib]{mn2e}

\usepackage{graphicx,subfigure}
\usepackage{bm}
\usepackage{aalongtable}
\usepackage{rotating}

\DeclareMathAlphabet{\mathsc}{OT1}{cmr}{m}{sc}
\def\testbx{bx}%
\DeclareRobustCommand{\ion}[2]{%
\relax\ifmmode
\ifx\testbx\f@series
{\mathbf{#1\,\mathsc{#2}}}\else
{\mathrm{#1\,\mathsc{#2}}}\fi
\else\textup{#1\,{\mdseries\textsc{#2}}}%
\fi}
\newcommand{\Hi} {\ion{H}{i}}

\newcommand{\ha} {\mbox{H$\alpha$}}
\newcommand{\hb} {\mbox{H$\beta$}}
\newcommand{\hg} {\mbox{H$\gamma$}}
\newcommand{\hd} {\mbox{H$\delta$}}
\newcommand{\Hei} {\ion{He}{i}}

\newcommand{\Ci} {\ion{C}{i}}
\newcommand{\Ni} {\ion{N}{i}}
\newcommand{\Nii} {\ion{N}{ii}}
\newcommand{\Oia} {[\ion{O}{i}]}
\newcommand{\Oi} {\ion{O}{i}}

\newcommand{\Nai} {\ion{Na}{i}}

\newcommand{\Siii} {\ion{Si}{ii}}

\newcommand{\Caiia} {[\ion{Ca}{ii}]}
\newcommand{\Caii} {\ion{Ca}{ii}}
\newcommand{\Scii} {\ion{Sc}{ii}}
\newcommand{\Tiii} {\ion{Ti}{ii}}

\newcommand{\Feiia} {[\ion{Fe}{ii}]}
\newcommand{\Feii} {\ion{Fe}{ii}}

\newcommand{\Baii} {\ion{Ba}{ii}}

\newcommand{\suspect}{SUSPECT}
\newcommand{\sn}{SN 2012aw}
\newcommand{\splot}{\texttt{SPLOT}}
\newcommand{\synow}{\texttt{SYNOW}}

\newcommand{\daomaster}{\texttt{DAOMASTER}}
\newcommand{\iraf}{\texttt{IRAF}}
\newcommand{\daophot}{\texttt{DAOPHOT}}
\newcommand{\lte}{\texttt{LTE}}
\newcommand{\nlte}{\texttt{NLTE}}
\newcommand{\cmfgen}{\texttt{CMFGEN}}
\newcommand{\phoneix}{\texttt{PHONEIX}}

\newcommand{\ebv}{\mbox{$E(B-V)$}}

\newcommand{\msun}{\mbox{M$_{\odot}$}}

\newcommand{\rsun}{\mbox{R$_{\odot}$}}
\newcommand{\kms}{\mbox{$\rm{\,km\,s^{-1}}$}}
\newcommand{\ergs}{\mbox{$\rm{\,erg\,s^{-1}}$}}

\newcommand{\nickel}{\mbox{$^{56}$Ni}}
\newcommand{\cobalt}{\mbox{$^{56}$Co}}
\newcommand{\iron}{\mbox{$^{56}$Fe}}
\newcommand{\mum}{\mbox{$\mu{\rm m}$}}
\newcommand{\el}{\mbox{${e}^{-}$}}

\begin{document}

\title[Optical follow-up observations of a type IIP supernova 2012aw]
{Supernova 2012aw - a high-energy clone of 
 archetypal type IIP SN 1999em}
\author[Bose et al.]
{Subhash Bose$^1$\thanks{e-mail: email@subhashbose.com, bose@aries.res.in}, 
 Brijesh Kumar$^1$, Firoza Sutaria$^2$, Brajesh Kumar$^{1,3}$, Rupak Roy$^1$,
\newauthor 
 V. K. Bhatt$^1$, S. B. Pandey$^1$, H. C. Chandola$^4$, Ram Sagar$^1$, Kuntal Misra$^1$,
 \newauthor
 Sayan Chakraborti$ ^5 $\\  
\newauthor
\\
 $^1$Aryabhatta Research Institute of Observational Sciences, Manora 
    Peak, Nainital - 263 002, India.\\
 $^2$Indian Institute of Astrophysics, Block-II, Koramangala, Bangalore - 560034, India.\\
 $^3$Institut d'Astrophysique et de G\'{e}ophysique, Universit\'{e} de 
 Li\`{e}ge, All\'{e}e du 6 Ao\^{u}t 17, B\^{a}t B5c, 4000 Li\`{e}ge, Belgium \\
 $^4$Department of Physics, D.S.B. Campus, Kumaun University, Nainital - 263 002, India.\\  
 $ ^5 $Institute for Theory and Computation, Harvard-Spithsonian Center for Astrophysics, 
       60 Garden Street, Cambridge, MA 02138, USA.\\
}

\date{Accepted 2013 May 10; Received 2013 April 27}

\pagerange{\pageref{firstpage}--\pageref{lastpage}} \pubyear{}

\maketitle

\label{firstpage}


\begin{abstract}
 We present densely-sampled $UBVRI$/$griz$ photometric and low-resolution (6-10\AA) 
 optical spectroscopic observations from 4 to 270 days after explosion of a newly 
 discovered type II {\sn} in a nearby ($\sim$9.9 Mpc) galaxy M95. 
 The light-curve characteristics of apparent magnitudes, colors, bolometric
 luminosity and the presence and evolution of prominent spectral features are found to 
 have striking similarity with the archetypal IIP SNe 1999em, 1999gi and 2004et. 
 The early time observations of \sn\ clearly detect minima in the 
 light-curve of $V$, $R$ and $I$ bands near 37 days after explosion and this we 
 suggest to be an observational evidence for emergence of recombination phase.
 The mid-plateau $M_{V}$ magnitude ($-16.67\pm0.04$) lies in between the bright ($\sim -18$) 
 and subluminous ($\sim -15$) IIP SNe. The mass of nickel is $0.06\pm0.01$ \msun.
 The \synow\ modelling of spectra indicate that the value and evolution of photospheric velocity   
 is similar to SN 2004et, but about $\sim$600 \kms\, higher than that of SNe 1999em and 1999gi
 at comparable epochs. This trend is more apparent in the line velocities of \ha\ and \hb.
 A comparison of ejecta velocity properties with that of existing radiation-hydrodynamical
 simulations indicate that the energy of explosion lies in the range 1-2$\times10^{51}$ ergs; 
 a further comparison of nebular phase \Oia\ doublet luminosity with SNe 2004et and 1987A 
 indicate that the mass of progenitor star is about 14-15 \msun. The presence of high-velocity 
 absorption features in the mid-to-late plateau and possibly in early phase spectra
 show signs of interaction between ejecta and the circumstellar matter; being consistent with 
 its early-time detection at X-ray and radio wavebands.
     
\end{abstract}	



\section{Introduction} \label{sec:intro}

 Early-time optical spectra of supernovae (SNe) showing strong Balmer lines of H are classified as 
 type II, whereas SNe I show no H lines. Many sub-types have been introduced \citep{1997ARA&A..35..309F} and 
 in IIP, the optical light curve remains constant for about hundred days (called the plateau phase) 
 and then decays exponentially, while both IIL and IIb are characterized by linear decline in the 
 light curves after reaching maxima at about twenty days after explosion \citep{2012ApJ...756L..30A}.
 The spectra of all these types show strong P-Cygni line profiles, while the type IIb has weak H 
 features initially and at later phases develop prominent He features similar to the type Ib SNe. 
 Type IIn events show narrow 
 width of H emission lines \citep{2002MNRAS.333...27P} which is indicative of interaction between 
 the SN ejecta and the dense circumstellar medium. In addition, there are peculiar type II events 
 such as SN 1987A which are characterized by long rise time of their light 
 curve \citep{2011A&A...532A.100U,2012A&A...537A.141P}. 

 Type II SNe are widely recognized as end stages of massive ($\ga$ 8 \msun) zero-age main-sequence
 stars which end up as core collapse explosions and having retained significant amount of H envelope 
 before explosion \citep{2013RvMP...85..245B}. The observed properties
 derived from the light-curve and spectra of SNe provide important clues to the understanding of
 explosion mechanisms as well as to the nature of progenitor stars \citep{2009MNRAS.395.1409S}.
 The plateau phase of IIP SNe is sustained by cooling down of the shock-heated expanding ejecta by 
 recombination of H while the post-plateau light curve is powered by the 
 radioactive decay of \cobalt\, into \iron, which in turn depend upon the amount of 
 \nickel\, synthesized during explosion. The observed properties of IIP SNe differ 
 greatly \citep{2003ApJ...582..905H,2009MNRAS.395.1409S}, 
 for example, the mid-plateau bolometric luminosity vary by an order of magnitude 
 from $\sim 10^{42.5}$\ergs\, for archetypal IIP SN 2004et to $\sim 10^{41.5}$\ergs\, for 
 under luminous SN 2005cs. The class of subluminous events \citep{2004MNRAS.347...74P,2009MNRAS.394.2266P} 
 are also accompanied by lower ejecta velocity during plateau ($\sim$ 1000 \kms) and lower luminosity 
 in the light curve tail owing to small yield of \nickel\, ($\sim 2-5\times10^{-3}$ \msun), in comparison 
 to the normal luminosity IIP SNe 1999em, 2004et with velocities ($\sim$ 5000 \kms) and the mass 
 of \nickel\, ($\sim$0.1 \msun). The subluminous IIP events are associated with O-Ne-Mg core 
 originating from lower-mass progenitors (8-10 \msun), while the normal ones originate from 
 iron-core collapse of massive ($>10\msun$) progenitors \citep{2011MNRAS.417.1417F,2012ARNPS..62..407J}. 
 However, there are cases, i.e. SN 2008in \citep{2011ApJ...736...76R} 
 and SN 2009js \citep{2013ApJ...767..166G}, where a spectroscopically subluminous IIP event 
 show light curve properties similar to a normal luminosity event.

 The stellar evolution models suggest that type IIP SNe originate from red supergiants having initial 
 masses between 9-25\msun\, having an upper mass cut of 32\msun\, for solar metallicity 
 stars \citep{2003ApJ...591..288H}, however, the observational constraints are 
 ambiguous and the mass of progenitors recovered from the analysis of pre-explosion archival HST images 
 for 20 IIP SNe lie in the range 9-17 \msun\, \citep{2009MNRAS.395.1409S} while the 
 hydrodynamical modelling of light curves for a handful of well studied IIP SNe indicate that they 
 primarily originate from 15-25 \msun\, progenitors \citep{2009A&A...506..829U,2011ApJ...729...61B}.
 Some of the problems in inferring physical properties are the lack of good quality
 data for nearby SNe. 


 SN 2012aw was discovered on March 16.9, 2012 by \cite{2012CBET.3054....1F} in the nearby 
 galaxy M95 ($\sim$ 10 Mpc) at R-band magnitude of 15. The first non-detection
 is reported on March 15.27 \citep{2012ATel.3996....1P} to a 3$\sigma$ limit of $R\sim20.7$. 
 Thus, we adopt March 16.1, 2012 (JD=2456002.6 $\pm$ 0.8 days) as the time of explosion (0 d) 
 through out the paper. The spectra obtained at 2d by \citet{2012CBET.3054....3M} showed a 
 featureless blue continuum while the subsequent spectra at later phases by  
 \citet{2012CBET.3054....2I,2012CBET.3054....4S} identifies the event 
 as young type IIP. The ultraviolet follow-up observations with UVOT/Swift is done 
 by \citet{2013ApJ...764L..13B} and similar to the optical, they report emergence of plateau
 in the UV light curve after 27d. The analysis of pre-explosion archival HST images of M95 in the 
 vicinity of \sn, by two independent group of researchers indicate that the progenitor was a 
 red superergiant with masses in the range 14-26\msun\,\citep{2012ApJ...759L..13F} 
 and 15-20\msun\,\citep{2012ApJ...756..131V} respectively.
 However, by accounting for appropriate extinction laws to the circumstellar and interstellar
 dust, \citet{2012ApJ...759...20K} determine that the luminosity of the progenitor star lie
 between 4.8 to 5.0 dex in solar units and the mass was less than 15\msun. \sn\, was
 also detected in X-rays observations with swift/XRT by \cite{2012ATel.3995....1I} and 
 in radio observations by \cite{2012ATel.4012....1S,2012ATel.4010....1Y} indicating
 interaction of ejecta with the circumstellar material. The early time (17d) optical
 spectropolarimetric observations with 8-m ESO VLT by \citet{2012ATel.4033....1L} finds 
 continuum polarization at the level of 0.3\% implying substantial asymmetries in the outer ejecta of \sn. 

 In this work, we present results from optical photometric ($UBVRI$ and/or $griz$) photometric 
 follow-up observations at 45 phases during 4d to 269d and low-resolution optical spectroscopic 
 observations at 14 phases during 7d to 270d of \sn. The paper is organized as follows. 
 In \S\ref{sec:obs.phot} and \S\ref{sec:obs.spec}, we present the photometric and spectroscopic 
 observations respectively and a brief description of light curves
 and spectra. Determination of reddening and extinction is given in \S\ref{sec:ext}. 
 In \S\ref{sec:lc}, we analyze light and color curves, derive bolometric light-curves and 
 estimate mass of nickel. In \S\ref{sec:sp}, we study spectra, evolution of spectral features, 
 \synow\ modelling and derive velocity of hydrogen envelope and the photosphere.
 The characteristics of explosion is described in \S\ref{sec:char} and conclusions are 
 presented in \S\ref{sec:sum}. 

 We adopt the distance to the host galaxy M95 as $9.9\pm0.1$ Mpc which is a weighted mean of 
 three most reliable redshift-independent distance measurements in the literature, i.e. $10\pm0.40$ Mpc
 by \citet{2001ApJ...553...47F} using cepheids; $9.86\pm0.14$ Mpc by \citet{2002ApJ...565..681R}
 using Tully-Fisher method and $9.83\pm0.13$ Mpc by \citet{2013IAUS..296..???T} using SN Expanding
 photosphere method. \sn\, occurred in the outskirts of the host galaxy at a deprojected distance
 of 6.8 kpc and the oxygen abundance (12+log[O/H]) of the galactic ISM at the position of SN is estimated
 as $8.8\pm0.1$ from the radial metallicity gradient relation in M95 given 
 by \citet{2006MNRAS.367.1139P} and this 
 value is close to the solar abundance for oxygen of 8.65 \citep{2009ARA&A..47..481A}. Some basic 
 properties of the host galaxy and \sn\, is listed in Table~\ref{tab:host}. 


  \begin{table}
  \caption{Properties of the host galaxy NGC 3351 and SN 2012aw.}
  \label{tab:host} 

  \begin{tabular}{llc} \hline \hline
     \noalign{\smallskip}
      Parameters& Value& Ref.$^{a}$\\ 
     \noalign{\smallskip} \hline
    
     \noalign{\smallskip}
     \multicolumn{3}{l}{\bf NGC 3351:}\\
     Type& Sb& 1\\
     RA (J2000)& $\alpha = 10^{\rm h} 43^{\rm m} 57\fs67$& 1\\
     DEC (J2000)& $\delta = 11\degr 42\arcmin 13\farcs0$& 1\\
     Abs. Magnitude& $M_{B}=-20.36$ mag& 1\\
     \\
     Distance& $D=9.9\pm0.1$ Mpc& \S\ref{sec:intro} \\
     Scale& $1\arcsec \sim 48$ pc, ~~$1\arcmin \sim 2.9$ kpc&\\
     Distance modulus& $\mu = 29.97\pm0.03$ &\\
     \\
     Apparent radius&${r_{\mathrm{25}}}=3\farcm6\,(\sim 10.5\,\mathrm{kpc})$& 1\\
     Inclination angle&$\Theta_{\rm inc}= 54.6^\circ$&1\\
     Position angle& $\Theta_{\rm maj}=9.9^\circ$&1\\
     Heliocentric Velocity& $cz_{\rm helio}=778\pm2 \kms$&1\\
     \\
     \multicolumn{3}{l}{\bf SN 2012aw:}\\
     RA (J2000)& $\alpha = 10^{\rm h} 43^{\rm m} 53\fs73$& 2\\
     DEC (J2000)& $\delta = 11\degr 40\arcmin 17\farcs9$& \\
     \\
     Galactocentric Location& 58\arcsec\,W, 115\arcsec\,S&  \\
     Deprojected radius&  $r_{\rm SN} = 139\farcs1$ ($\sim$ 6.75 kpc)& \\
     \\
     Time of explosion & $t_{\rm 0}$ =16.1 March 2012 (UT)& \S\ref{sec:intro}\\ 
                    & (JD 2456002.59)& \\ 
     Reddening & \ebv\,=$0.074\pm0.008$& \S\ref{sec:ext}\\
   
     \noalign{\smallskip}
     \hline
  \end{tabular}
  \newline $^{a}$ (1) HyperLEDA - http://leda.univ-lyon1.fr;\\
                  (2) \citet{2012ApJ...756..131V}
  \end{table}


\section{Observation and data reduction} \label{sec:obs}

\subsection{Photometry} \label{sec:obs.phot}

 The broadband photometric data in $UBVRI$ Johnson-Cousins and $griz$ SDSS systems are collected 
 using the 104-cm Sampurnanand Telescope (ST) at Manora Peak, Nainital and the 130-cm Devasthal 
 Fast Optical Telescope (DFOT) at Devasthal, Nainital. Both the telescopes are operated by the 
 Aryabhatta Research Institute of Observational sciences, India \citep{2013arXiv1304.0235S}. The 104-cm ST 
 is equipped with a 2k$\times$2k liquid-nitrogen cooled CCD camera having square pixels of 24\,\mum\,; 
 and with a plate scale of 0\farcs37 per pixel, the CCD covers a square field-of-view of 
 about 13\arcmin\, on a side in the sky. While operating at 27 kHz, the gain and readout noise of 
 the CCD are 10\el\, per analog-to-digital unit and 5.3\el\, respectively. The 130-cm DFOT is 
 equipped with a 2k$ \times $2k Peltier-cooled CCD camera having a pixel size of 13.5\,\mum\,; and with a plate
 scale of 0\farcs54 per pixel, the CCD covers a square field-of-view of 18\arcmin\, on a side. The CCD was
 operated at 31 kHz speed with readout noise of 2.5\el\, -- a detailed technical 
 description and performance of this camera can be found elsewhere \citep{2012ASInC...4..173S}.   
 A binning of 2$\times$2 is used in CCDs wherever required to improve the signal-to-noise ratio. 
 At 104-cm ST, we had $UBVRI$ while at 130-cm, we had $BVR$ as well as $griz$ filters. A typical exposure 
 time of 300s were given for $U$ and $B$ filters while 80-200s were given for remaining filters. 
 For V-band, the full width at half maximum (FWHM) of the stellar point spread function (PSF) varied 
 between 2\farcs1 to 3\farcs5, with a median value of 2\farcs6. In addition to the target exposures, 
 several bias and twilight flat frames are also obtained for calibration purpose.

\begin{figure}
\centering
\includegraphics[bb=0.0 0.0 123.0 121.68,width=8.4cm]{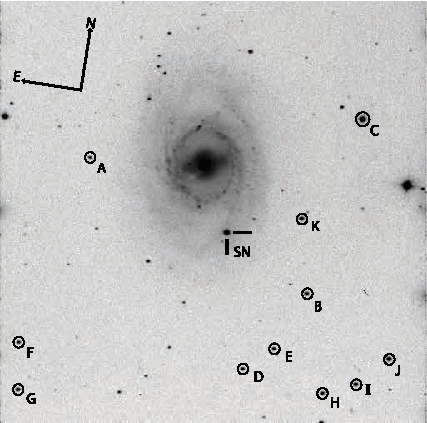}
\caption{\sn\ in NGC 3551. The $V$-band image taken from 104-cm ST covering an area of 
         about 13\arcmin$\times$13\arcmin\, is shown. The location of SN and local 
         standard stars are marked.} 
\label{fig:snid}
\end{figure}

 The bias subtraction, flat fielding, cosmic ray removal, alignment and determination 
 of mean FWHM and ellipticity for each object frames are done using the standard tasks 
 available in the data reduction softwares {\iraf}\footnote{{\iraf} stands for 
 Image Reduction and Analysis Facility distributed by the National Optical Astronomy Observatories which
 is operated by the Association of Universities for research in Astronomy,
 Inc. under co-operative agreement with the National Science Foundation.}
 and {\daophot}\,\footnote{ {\daophot} stands for Dominion Astrophysical 
 Observatory Photometry \citep{1987PASP...99..191S}}.
 Whenever multiple frames are available, the photometry is performed on co-added frames. 
 As the location of SN is fairly isolated from the galaxy center and it lies on a smooth and 
 faint galaxy background (see Fig.~\ref{fig:snid}), we chose to perform aperture photometry; and an 
 aperture radius equal to the mean FWHM of a given frame and a 10 pixel wide sky annulus at 6 FWHM 
 radius were chosen. This whole scheme of aperture photometry is achieved using 
 standalone version of \daophot\ subroutines.
 The differential instrumental magnitude for each filter was generated using \daomaster\ task.

\begin{table*}
  \caption{Identification number (ID), coordinates ($\alpha, \delta$) and calibrated
           magnitudes of stable secondary standard stars in the field of SN 2012aw. The
           field of {\sn} was calibrated using Landolt standards observed on the night of
           22nd March 2012. The quoted errors in magnitude include both photometric and calibration
           errors and it denote 1$\sigma$ uncertainty.}
  \label{tab:photstar}

  \begin{tabular}{cccccccc}
     \hline
     Star& $\alpha_{\rm J2000}$&       $\delta_{\rm J2000}$& $U$& $B$&  $V$&  $R$&  $I$\\
       ID&          (h m s)&(\degr\,\arcmin\,\arcsec)&(mag)&(mag)&(mag)&(mag)&(mag)\\
     \hline
    A& 10:44:11.82& +11:41:54.5&   18.168 $\pm$ 0.078 & 17.650 $\pm$ 0.045 & 16.775 $\pm$ 0.057 & 16.316 $\pm$ 0.024 & 15.841 $\pm$ 0.017\\
    B& 10:43:42.70& +11:38:51.1&   15.801 $\pm$ 0.024 & 15.611 $\pm$ 0.008 & 14.897 $\pm$ 0.011 & 14.493 $\pm$ 0.013 & 14.049 $\pm$ 0.016\\
    C& 10:43:39.17& +11:44:21.3&   13.774 $\pm$ 0.014 & 12.882 $\pm$ 0.007 & 11.878 $\pm$ 0.026 & 11.358 $\pm$ 0.035 & 10.866 $\pm$ 0.035\\
    D& 10:43:49.09& +11:36:17.4&   16.277 $\pm$ 0.024 & 16.179 $\pm$ 0.012 & 15.522 $\pm$ 0.017 & 15.139 $\pm$ 0.021 & 14.748 $\pm$ 0.019\\
    E& 10:43:45.65& +11:37:02.8&   16.804 $\pm$ 0.037 & 16.453 $\pm$ 0.012 & 15.668 $\pm$ 0.013 & 15.243 $\pm$ 0.016 & 14.817 $\pm$ 0.022\\
    F& 10:44:17.02& +11:36:02.1&   16.540 $\pm$ 0.026 & 16.760 $\pm$ 0.012 & 16.399 $\pm$ 0.014 & 16.123 $\pm$ 0.026 & 15.795 $\pm$ 0.016\\
    G& 10:44:16.18& +11:34:37.0&   16.277 $\pm$ 0.031 & 15.687 $\pm$ 0.015 & 14.851 $\pm$ 0.014 & 14.397 $\pm$ 0.019 & 13.966 $\pm$ 0.018\\
    H& 10:43:38.86& +11:35:56.5&   16.160 $\pm$ 0.023 & 15.919 $\pm$ 0.012 & 15.207 $\pm$ 0.016 & 14.818 $\pm$ 0.020 & 14.453 $\pm$ 0.023\\
    I& 10:43:34.96& +11:36:22.2&   15.746 $\pm$ 0.022 & 15.494 $\pm$ 0.009 & 14.787 $\pm$ 0.012 & 14.406 $\pm$ 0.025 & 13.966 $\pm$ 0.016\\
    J& 10:43:31.33& +11:37:17.7&   15.954 $\pm$ 0.018 & 14.955 $\pm$ 0.006 & 13.896 $\pm$ 0.013 & 13.300 $\pm$ 0.025 & 12.712 $\pm$ 0.016\\
    K& 10:43:44.75& +11:41:04.4&   15.190 $\pm$ 0.020 & 15.328 $\pm$ 0.009 & 14.944 $\pm$ 0.009 & 14.727 $\pm$ 0.012 & 14.467 $\pm$ 0.014\\
     \hline
  \end{tabular}

\end{table*}

\begin{table*}
  \fontsize{2mm}{2mm}\selectfont
  \caption{Photometric evolution of \sn. Errors denote $1\sigma$ uncertainty.}
  \label{tab:photsn}
  \begin{tabular}
  {c c c c c c c c l c}
  \hline
  UT Date&JD&Phase$^{a}$&$U$&$B$&$V$&$R$&$I$&Tel$^{b}$& Seeing$^{c}$ \\
  (yy/mm/dd)&2456000+&(day)&(mag)&(mag)&(mag)&(mag)&(mag)&& (\arcsec)\\
  \hline
2012-03-19.82 &  006.32 &   4 &  12.671  $\pm$   0.042  &  13.504  $\pm$   0.021  &  13.574  $\pm$   0.017  &  13.463  $\pm$   0.018  &  13.433  $\pm$   0.025  &  ST   &  2.6 \\
2012-03-20.80 &  007.30 &   5 &  12.582  $\pm$   0.042  &  13.492  $\pm$   0.021  &  13.478  $\pm$   0.017  &  13.372  $\pm$   0.018  &  13.325  $\pm$   0.025  &  ST   &  2.3 \\
2012-03-21.81 &  008.31 &   6 &  12.517  $\pm$   0.042  &  13.457  $\pm$   0.021  &  13.420  $\pm$   0.012  &  13.300  $\pm$   0.018  &  13.258  $\pm$   0.018  &  ST   &  2.0 \\
2012-03-22.76 &  009.26 &   7 &  12.599  $\pm$   0.042  &  13.415  $\pm$   0.021  &  13.380  $\pm$   0.017  &  13.262  $\pm$   0.018  &  13.239  $\pm$   0.025  &  ST   &  2.0 \\
2012-03-24.84 &  011.34 &   9 &  12.652  $\pm$   0.042  &  13.394  $\pm$   0.021  &  13.340  $\pm$   0.017  &  13.175  $\pm$   0.018  &  13.132  $\pm$   0.025  &  ST   &  1.9 \\
2012-03-25.78 &  012.28 &  10 &  12.676  $\pm$   0.043  &  13.411  $\pm$   0.021  &  13.328  $\pm$   0.017  &  13.192  $\pm$   0.014  &  13.130  $\pm$   0.025  &  ST   &  2.3 \\
2012-04-01.63 &  019.13 &  17 &  13.010  $\pm$   0.042  &  13.567  $\pm$   0.021  &  13.318  $\pm$   0.017  &  13.089  $\pm$   0.018  &  12.990  $\pm$   0.025  &  ST   &  2.2 \\
2012-04-07.63 &  025.13 &  23 &  13.627  $\pm$   0.043  &  13.756  $\pm$   0.021  &  13.325  $\pm$   0.017  &  13.073  $\pm$   0.018  &  12.923  $\pm$   0.025  &  ST   &  2.2 \\
2012-04-09.74 &  027.24 &  25 &  13.909  $\pm$   0.042  &  13.875  $\pm$   0.021  &  13.351  $\pm$   0.017  &  13.086  $\pm$   0.018  &  12.950  $\pm$   0.025  &  ST   &  2.3 \\
2012-04-12.71 &  030.21 &  28 &  14.199  $\pm$   0.042  &  14.000  $\pm$   0.021  &  13.366  $\pm$   0.017  &  13.111  $\pm$   0.018  &  12.940  $\pm$   0.025  &  ST   &  2.4 \\
2012-04-19.65 &  037.15 &  35 &  14.788  $\pm$   0.043  &  14.228  $\pm$   0.021  &  13.439  $\pm$   0.017  &  13.140  $\pm$   0.018  &  12.918  $\pm$   0.025  &  ST   &  2.5 \\
2012-04-21.61 &  039.11 &  37 &  14.905  $\pm$   0.043  &  14.284  $\pm$   0.021  &  13.468  $\pm$   0.017  &  13.151  $\pm$   0.018  &  12.895  $\pm$   0.018  &  ST   &  2.9 \\
2012-04-22.82 &  040.32 &  38 &           ---           &           ---           &  13.493  $\pm$   0.017  &  13.135  $\pm$   0.018  &  12.899  $\pm$   0.052  &  ST   &  2.7 \\
2012-04-25.75 &  043.25 &  41 &  15.189  $\pm$   0.044  &  14.392  $\pm$   0.021  &  13.463  $\pm$   0.017  &  13.124  $\pm$   0.018  &  12.851  $\pm$   0.025  &  ST   &  2.1 \\
2012-04-29.70 &  047.20 &  45 &  15.377  $\pm$   0.050  &  14.466  $\pm$   0.021  &  13.487  $\pm$   0.017  &  13.126  $\pm$   0.018  &  12.836  $\pm$   0.025  &  ST   &  2.8 \\
2012-05-01.61 &  049.11 &  47 &  15.455  $\pm$   0.057  &  14.491  $\pm$   0.025  &  13.725  $\pm$   0.024  &  13.115  $\pm$   0.023  &  12.887  $\pm$   0.035  &  ST   &  2.9 \\
2012-05-04.63 &  052.13 &  50 &  15.552  $\pm$   0.089  &           ---           &  13.439  $\pm$   0.018  &  13.115  $\pm$   0.018  &  12.776  $\pm$   0.055  &  ST   &  2.1 \\
2012-05-06.62 &  054.12 &  52 &  15.725  $\pm$   0.041  &  14.574  $\pm$   0.021  &  13.471  $\pm$   0.017  &  13.116  $\pm$   0.018  &  12.790  $\pm$   0.052  &  ST   &  2.4 \\
2012-05-08.61 &  056.11 &  54 &           ---           &           ---           &  13.526  $\pm$   0.018  &  13.113  $\pm$   0.019  &  12.812  $\pm$   0.026  &  ST   &  2.7 \\
2012-05-09.61 &  057.11 &  55 &  15.802  $\pm$   0.067  &  14.688  $\pm$   0.024  &  13.494  $\pm$   0.017  &  13.088  $\pm$   0.018  &  12.775  $\pm$   0.025  &  ST   &  2.9 \\
2012-05-11.65 &  059.15 &  57 &           ---           &  14.518  $\pm$   0.022  &  13.534  $\pm$   0.017  &  13.085  $\pm$   0.018  &           ---           &  DFOT &  3.1 \\
2012-05-13.68 &  061.18 &  59 &  16.049  $\pm$   0.059  &  14.697  $\pm$   0.021  &  13.539  $\pm$   0.017  &  13.097  $\pm$   0.018  &  12.804  $\pm$   0.025  &  ST   &  2.6 \\
2012-05-14.69 &  062.19 &  60 &  16.133  $\pm$   0.068  &  14.711  $\pm$   0.022  &  13.566  $\pm$   0.017  &  13.139  $\pm$   0.018  &  12.797  $\pm$   0.025  &  ST   &  2.6 \\
2012-05-17.64 &  065.14 &  63 &  16.239  $\pm$   0.049  &  14.771  $\pm$   0.021  &  13.556  $\pm$   0.017  &  13.102  $\pm$   0.018  &  12.774  $\pm$   0.025  &  ST   &  2.8 \\
2012-05-24.63 &  072.13 &  70 &  16.424  $\pm$   0.089  &  14.817  $\pm$   0.021  &  13.571  $\pm$   0.017  &  13.106  $\pm$   0.018  &  12.787  $\pm$   0.025  &  ST   &  3.1 \\
2012-05-27.66 &  075.16 &  73 &  16.577  $\pm$   0.144  &  14.882  $\pm$   0.022  &  13.596  $\pm$   0.018  &  13.118  $\pm$   0.018  &  12.778  $\pm$   0.025  &  ST   &  2.9 \\
2012-05-27.65 &  075.15 &  73 &           ---           &  14.611  $\pm$   0.023  &  13.596  $\pm$   0.012  &  13.110  $\pm$   0.018  &           ---           &  DFOT &  2.8 \\
2012-05-28.63 &  076.13 &  74 &           ---           &  14.609  $\pm$   0.025  &  13.637  $\pm$   0.018  &  13.115  $\pm$   0.018  &           ---           &  DFOT &  3.1 \\
2012-05-30.65 &  078.15 &  76 &           ---           &  14.872  $\pm$   0.031  &  13.627  $\pm$   0.018  &  13.126  $\pm$   0.018  &  12.786  $\pm$   0.025  &  ST   &  3.8 \\
2012-06-06.64 &  085.14 &  83 &  16.840  $\pm$   0.061  &  14.990  $\pm$   0.021  &  13.687  $\pm$   0.017  &  13.127  $\pm$   0.018  &  12.804  $\pm$   0.018  &  ST   &  3.4 \\
2012-06-09.62 &  088.12 &  86 &  17.010  $\pm$   0.062  &  15.042  $\pm$   0.022  &  13.664  $\pm$   0.017  &  13.174  $\pm$   0.018  &  12.828  $\pm$   0.025  &  ST   &  2.5 \\
2012-06-11.63 &  090.13 &  88 &  16.886  $\pm$   0.052  &  15.057  $\pm$   0.021  &  13.670  $\pm$   0.017  &  13.159  $\pm$   0.018  &  12.815  $\pm$   0.025  &  ST   &  2.8 \\
2012-06-16.66 &  095.16 &  93 &           ---           &  14.694  $\pm$   0.029  &  13.727  $\pm$   0.018  &  13.206  $\pm$   0.013  &           ---           &  DFOT &  3.3 \\
2012-06-25.65 &  104.15 & 102 &           ---           &  15.231  $\pm$   0.029  &  13.873  $\pm$   0.022  &  13.361  $\pm$   0.023  &  12.995  $\pm$   0.031  &  ST   &  3.7 \\
2012-06-27.63 &  106.13 & 104 &  16.979  $\pm$   0.219  &  15.158  $\pm$   0.018  &  13.824  $\pm$   0.017  &  13.296  $\pm$   0.018  &  12.947  $\pm$   0.025  &  ST   &  3.7 \\
2012-10-16.96 &  217.46 & 215 &           ---           &  17.526  $\pm$   0.083  &  16.577  $\pm$   0.025  &  15.794  $\pm$   0.020  &           ---           &  DFOT &  3.5 \\
2012-10-17.96 &  218.46 & 216 &           ---           &  17.657  $\pm$   0.053  &  16.617  $\pm$   0.024  &  15.823  $\pm$   0.014  &           ---           &  DFOT &  3.5 \\
2012-10-23.98 &  224.48 & 222 &           ---           &  18.069  $\pm$   0.361  &  16.918  $\pm$   0.366  &           ---           &  15.413  $\pm$   0.059  &  ST   &  3.4 \\
2012-10-24.97 &  225.47 & 223 &           ---           &  18.033  $\pm$   0.026  &  16.711  $\pm$   0.019  &  15.832  $\pm$   0.019  &  15.368  $\pm$   0.026  &  ST   &  2.2 \\
2012-10-26.97 &  227.47 & 225 &           ---           &  18.106  $\pm$   0.023  &  16.730  $\pm$   0.019  &  15.860  $\pm$   0.019  &  15.389  $\pm$   0.026  &  ST   &  2.7 \\
2012-10-29.99 &  230.49 & 228 &           ---           &           ---           &  16.667  $\pm$   0.015  &           ---           &           ---           &  ST   &  1.7 \\
2012-10-30.98 &  231.48 & 229 &  19.551  $\pm$   0.361  &  18.088  $\pm$   0.038  &  16.765  $\pm$   0.020  &  15.899  $\pm$   0.014  &  15.447  $\pm$   0.025  &  ST   &  1.8 \\
2012-11-02.97 &  234.47 & 232 &           ---           &  18.163  $\pm$   0.030  &  16.803  $\pm$   0.020  &  15.949  $\pm$   0.011  &  15.454  $\pm$   0.026  &  ST   &  2.0 \\
2012-11-04.96 &  236.46 & 234 &           ---           &  18.126  $\pm$   0.030  &  16.841  $\pm$   0.016  &  15.975  $\pm$   0.020  &  15.456  $\pm$   0.026  &  ST   &  2.0 \\
2012-11-07.97 &  239.47 & 237 &           ---           &  18.278  $\pm$   0.078  &  16.721  $\pm$   0.046  &  15.955  $\pm$   0.031  &  15.454  $\pm$   0.040  &  ST   &  2.2 \\
2012-11-10.95 &  242.45 & 240 &           ---           &  17.850  $\pm$   0.051  &  16.911  $\pm$   0.022  &  15.992  $\pm$   0.019  &           ---           &  DFOT &  2.9 \\
2012-11-11.94 &  243.44 & 241 &           ---           &  17.875  $\pm$   0.051  &  16.963  $\pm$   0.021  &  15.996  $\pm$   0.019  &           ---           &  DFOT &  3.0 \\
2012-11-16.98 &  248.48 & 246 &  20.291  $\pm$   0.239  &  18.264  $\pm$   0.019  &  16.908  $\pm$   0.014  &  16.039  $\pm$   0.019  &  15.586  $\pm$   0.019  &  ST   &  2.0 \\
2012-11-18.98 &  250.48 & 248 &  19.934  $\pm$   0.153  &  18.289  $\pm$   0.020  &  16.927  $\pm$   0.018  &  16.054  $\pm$   0.019  &  15.610  $\pm$   0.026  &  ST   &  1.7 \\
2012-11-25.95 &  257.45 & 255 &  19.821  $\pm$   0.111  &  18.323  $\pm$   0.018  &  17.011  $\pm$   0.014  &  16.132  $\pm$   0.019  &  15.709  $\pm$   0.026  &  ST   &  1.9 \\
2012-12-04.90 &  266.40 & 264 &           ---           &  18.325  $\pm$   0.060  &  17.051  $\pm$   0.040  &  16.012  $\pm$   0.036  &  15.726  $\pm$   0.050  &  ST   &  2.0 \\
2012-12-04.86 &  266.36 & 264 &           ---           &           ---           &  17.130  $\pm$   0.071  &           ---           &           ---           &  DFOT &  3.0 \\
2012-12-09.02 &  270.52 & 268 &           ---           &  18.415  $\pm$   0.070  &  17.109  $\pm$   0.022  &  16.225  $\pm$   0.020  &  15.811  $\pm$   0.026  &  ST   &  2.6 \\
2012-12-09.98 &  271.48 & 269 &  20.166  $\pm$   0.165  &  18.429  $\pm$   0.022  &  17.113  $\pm$   0.019  &  16.242  $\pm$   0.019  &  15.837  $\pm$   0.026  &  ST   &  2.2 \\  
    \hline
    & & & $g$& $r$& $i$& $z$& & & \\                                                              
    & & &(mag)& (mag)& (mag)& (mag)& & & \\ \hline                                                     
2012-05-11.66 &  059.16 &   57 &  13.922 $\pm$ 0.031     &  13.242 $\pm$ 0.028     & 13.187 $\pm$ 0.021      & 13.185 $\pm$ 0.070      &                         &  DFOT &  3.1  \\
2012-05-27.68 &  075.18 &   73 &  14.038 $\pm$ 0.031     &  13.259 $\pm$ 0.028     & 13.199 $\pm$ 0.021      & 13.164 $\pm$ 0.049      &                         &  DFOT &  2.8  \\
2012-05-28.65 &  076.15 &   74 &  14.052 $\pm$ 0.031     &  13.269 $\pm$ 0.028     & 13.192 $\pm$ 0.021      & 13.172 $\pm$ 0.070      &                         &  DFOT &  3.1  \\
2012-06-16.67 &  095.17 &   93 &  14.173 $\pm$ 0.031     &  13.326 $\pm$ 0.028     & 13.259 $\pm$ 0.021      & 13.192 $\pm$ 0.070      &                         &  DFOT &  3.3  \\
2012-10-16.97 &  217.47 &  215 &  17.150 $\pm$ 0.033     &  15.948 $\pm$ 0.028     & 15.907 $\pm$ 0.022      & 15.376 $\pm$ 0.071      &                         &  DFOT &  3.5  \\
2012-10-17.98 &  218.48 &  216 &  17.161 $\pm$ 0.033     &  15.971 $\pm$ 0.028     & 15.914 $\pm$ 0.022      & 15.399 $\pm$ 0.071      &                         &  DFOT &  3.5  \\
2012-11-10.93 &  242.43 &  240 &  17.339 $\pm$ 0.033     &  16.171 $\pm$ 0.028     & 16.158 $\pm$ 0.022      & 15.570 $\pm$ 0.071      &                         &  DFOT &  2.9  \\
2012-11-11.93 &  243.43 &  241 &  17.353 $\pm$ 0.033     &  16.173 $\pm$ 0.028     & 16.167 $\pm$ 0.022      &         ---             &                         &  DFOT &  3.0  \\
2012-12-04.95 &  266.45 &  264 &          ---            &          ---            &         ---             &         ---             &                         &  DFOT &  3.0  \\
  \hline                                                                                                          
  \end{tabular}
\begin{flushleft}
  $^{a}$ with reference to the explosion epoch JD 2456002.59\\
  $^{b}$ ST : 104-cm Sampurnanand Telescope, ARIES, India; DFOT : 130-cm Devasthal fast optical telescope, ARIES, India\\
  $^{c}$ FWHM of the median stellar PSF at $V$ band frame.\\
\end{flushleft}
\end{table*}

 In order to calibrate instrumental magnitude of \sn\,, the \citet{2009AJ....137.4186L} 
 standard fields PG~1323, PG~1525 and PG~1633 were observed on 22nd March 2012 in $UBVRI$ filters with 
 the 104-cm ST under photometric night conditions viz. transparent sky, FWHM seeing 
 in $V \sim 2\arcsec$. The observations of standard fields were taken at five locations covering  
 airmass from 1.06 to 2.36. The SN field is also observed on the same night. The data 
 reduction of SN and Landolt fields are done using profile fitting technique and the 
 instrumental magnitudes were converted into 
 standard system following least-square linear regression procedures outlined 
 in \citet{1992JRASC..86...71S}, in which zero-points, color coefficients and the 
 atmospheric extinction coefficients are fitted simultaneously for 17 stars having 
 V magnitude from 12.0 to 16.4 and $B-V$ from $-0.22$ to 1.14 mag. 
%
%
 The root-mean-squared (RMS) scatter between the transformed and the standard magnitudes of Landolt stars is 
 found to be $\sim$ 0.04 mag in $U$, 0.02 mag in $B$ and 0.01 mag in $VRI$. The transformation coefficients 
 were used to generate local standard stars in the field of \sn\, observed on the same night and 
 a set of 11 stars with $V$ range from 12 to 17 mag and $B-V$ range of 0.36 to 1.06 mag were selected.
 These stars are identified in Fig.~\ref{fig:snid} and are listed in Table~\ref{tab:photstar}. The quoted 
 errors include both photometric and calibration errors propagated in quadrature. The frame-to-frame
 photometric variability of these stars during 4d to 269d were found to lie within quoted uncertainties. 
 Four of these stars are in common with the study of \citet{2012IBVS.6024....1H} who provided calibrated 
 $BVRI$ magnitudes of 14 stars in 20\arcmin $\times$ 20\arcmin\, field of \sn. A comparison 
 with their photometry gives a mean and RMS scatter of $-0.04\pm0.04$, $-0.01\pm0.03$, 
 $0.02\pm0.04$ and $0.00\pm0.03$ mag respectively for $B$, $V$, $R$ and $I$; indicating that the two
 photometric measurements are consistent within uncertainties. The magnitude of local
 standards in $griz$ were taken from Lupton et al. 2005.
 The final photometry of \sn\, is given in Table~\ref{tab:photsn}. 


\subsection{Spectroscopy} \label{sec:obs.spec} 

%
%
%
%
%
%
%

 During 7d to 270d, long-slit low-resolution spectra in the optical range were collected at 14 epochs; 
 nine from 2m IUCAA Girawali Observatory (IGO) telescope and five from 2m Himalayan Chandra 
 Telescope (HCT). Journal of spectroscopic observations are given in Table~\ref{tab:speclog}. 
 Observations at 2m IGO have been carried out using IUCAA Faint Object Spectrograph and 
 Camera (IFOSC) mounted at the cassegrain focus of f/10 
 reflector \citep{2002BASI..30...785,2005BASI...33..513C}. 
 Grisms 5 ($\lambda\sim$  0.33-0.63 \mum; $\Delta\lambda\sim$8.8 \AA) and pair of 
 grism 7 \& 8 ($\lambda\sim$  0.38-0.83\mum; $\Delta\lambda\sim$4 \AA) along with a slit width of 1\farcs5 
 are used to record spectra on $2048\times2048$ E2V CCD with 13.5 \mum\,pixel size; having a gain 
 of 1.5 \el\, per analog-to-digital unit (ADU) and readout noise of 4 \el. Calibration frames (bias, flat, 
 HeNe arc) and spectrophotometric flux standard stars (Feige34, Hr4468 and Hz44) were observed on each night. 
 Observations from 2m HCT have been carried out in similar fashion with HFOSC using pair of 
 grisms 7\&8 ($\lambda\sim$  0.38-0.84\mum; $\Delta\lambda\sim$4 \AA) and a slit width of 1\farcs92.
 For flux calibration, stars Feige66, Feige110 and Grw70d5824 are observed and FeAr and FeNe arcs 
 are observed for wavelength calibrations.   
 
 Spectroscopic data reduction was done under \iraf\ environment. Bias and flat fielding were
 performed on each frames. Cosmic ray rejection on each frame were done using 
 Laplacian kernel detection algorithm for spectra, L.A.Cosmic \citep{2001PASP..113.1420V}.
 One-dimensional spectra were extracted using {\it apall} task which is based on
 optimal extraction algorithm by \citet{1986PASP...98..609H}. Wavelength calibration were performed 
 using {\it identify} task and about 15-18 emission lines of HeNe (for IGO) or FeNe and 
 FeAr (for HCT) were used to find dispersion solution.
 The position of OI emission skyline at 5577\AA\, was used to check the wavelength 
 calibration and deviations were found to lie between 0.3 to 5.5\AA\, and this was corrected by 
 applying a linear shift in wavelength.
 The instrumental FWHM resolution of 2m IGO spectra as 
 measured from \Oi\,5577\AA\, emission skyline was found to lie between $\sim$6 to 12\AA\, 
 ($\sim$ 317 - 671 \kms) and that of 2m HCT was found to lie between $\sim$8 to 10\AA\,.  

 The flux calibration of wavelength-calibrated spectra was done using \textit{standard, sensfunc 
 and calibrate} tasks. We used
 spectrophotometric standard fluxes from \citet{1990AJ.....99.1621O,1994PASP..106..566H} and 
 the spectral extinction coefficients from \citet{2008BASI...36..111S,2005BASI...33..513C} for the 
 respective sites.
 All the spectra were tied to an absolute flux scale using zeropoints determined from $UBVRI$ magnitudes. 
 To tie the spectra with photometry, individual spectrum is multiplied by wavelength 
 dependent polynomial function and it's $ BVRI $ filter-response convolved fluxes are compared 
 with photometric fluxes at corresponding epoch. The multiplied polynomial is tuned to minimize 
 the flux difference and obtain the tied spectrum. 
 The one-dimensional spectra are corrected for 
 heliocentric velocity of the 
 host galaxy (778 \kms;\S\ref{sec:intro}) using {\it dopcor} tasks.


\begin{table*}
  \caption{Journal of optical spectroscopic observations of \sn. The spectral observations are made at 14 
           phases between 7d to 270d.}
  \label{tab:speclog}
  \begin{tabular}{lc cc cc cc cr}
    \hline
    UT Date &JD&Phase$^{a}$&    Range$^{b}$&Telescope$^{c}$& Grating& Slit width& Dispersion& Exposure&S/N$^{d}$ \\
    (yy/mm/dd.dd) &2456000+&(days)& \mum&      & (gr mm$^{-1}$)&  (\arcsec)& (\AA\,pix$^{-1}$)& (s)&(pix$^{-1}$)\\ \hline

        2012-03-22.752&  9.25&  7& 0.38$-$0.68& HCT& 600& 1.92&1.43&   900     &  85   \\
                      &      &   & 0.58$-$0.84& HCT& 600& 1.92&1.26&   900     &  66   \\
        2012-03-23.768& 10.26&  8& 0.33$-$0.63& IGO& 300& 1.50&2.75&   900     & 147   \\
        2012-03-27.615& 14.12& 12& 0.38$-$0.68& HCT& 600& 1.92&1.49&  1200     & 123   \\
                      &      &   & 0.58$-$0.84& HCT& 600& 1.92&1.26&   900     &  75   \\
        2012-03-30.659& 17.16& 15& 0.38$-$0.68& HCT& 600& 1.92&1.49&   900     &  91   \\
                      &      &   & 0.58$-$0.84& HCT& 600& 1.92&1.26&   900     &  68   \\
        2012-03-31.809& 18.30& 16& 0.38$-$0.68& HCT& 600& 1.92&1.49&   900     &  48   \\
                      &      &   & 0.58$-$0.84& HCT& 600& 1.92&1.26&   900     &  35   \\
        2012-04-04.714& 22.21& 20& 0.38$-$0.68& HCT& 600& 1.92&1.61&  1200     & 127   \\
                      &      &   & 0.58$-$0.84& HCT& 600& 1.92&1.25&  1200     &  86   \\
        2012-04-10.715& 28.21& 26& 0.38$-$0.68& IGO& 600& 1.50&1.39&  2700     & 112   \\
                      &      &   & 0.58$-$0.83& IGO& 600& 1.50&1.16&  1200,2104&  71   \\
        2012-04-15.802& 33.30& 31& 0.33$-$0.63& IGO& 300& 1.50&2.89&  1800x2   & 171   \\
        2012-04-29.685& 47.18& 45& 0.33$-$0.63& IGO& 300& 1.50&2.89&  1200x2   & 114   \\
        2012-05-09.713& 57.22& 55& 0.38$-$0.68& HCT& 600& 1.92&1.61&  1200     & 139   \\
                      &      &   & 0.58$-$0.84& HCT& 600& 1.92&1.25&  1200     & 104   \\
        2012-05-15.680& 63.19& 61& 0.38$-$0.68& IGO& 600& 1.50&1.38&  1800     & 103   \\
                      &      &   & 0.58$-$0.83& IGO& 600& 1.50&1.16&  1800     &  85   \\
	    2012-05-20.730& 68.23& 66& 0.38$-$0.68& HCT& 600& 1.92&1.61&   900     &  95   \\
                      &      &   & 0.58$-$0.84& HCT& 600& 1.92&1.25&   900     &  64   \\
	    2012-06-27.629&106.13&104& 0.38$-$0.68& HCT& 600& 1.92&1.61&  1200     &  56   \\
                      &      &   & 0.58$-$0.84& HCT& 600& 1.92&1.25&  1200     &  38   \\
	    2012-12-10.846&272.37&270& 0.38$-$0.68& HCT& 600& 1.92&1.43&  2400     &  26   \\
                      &      &   & 0.58$-$0.84& HCT& 600& 1.92&1.26&  2400     &  20   \\
    \hline
  \end{tabular}
\begin{flushleft}
  $^{a}$ With reference to the burst time JD 2456002.59\\
  $^{b}$ For transmission $\ge$50\%\\
  $^{c}$ HCT : HFOSC on 2 m Himalyan Chandra Telescope, Hanle; IGO : IFOSC on
  2 m IUCAA Girawali Observatory, India. \\
  $^{d}$ At 0.6 \mum\\
\end{flushleft}
\end{table*}

\section{Extinction} \label{sec:ext} 

%
%
%
%
%

 In order to derive intrinsic properties of explosion, the reddening due to interstellar matter
 in both the Milky Way and the host galaxy, towards the sight-line of \sn\ should be known 
 accurately. Using all-sky dust-extinction map of \cite{1998ApJ...500..525S}, we derived 
 the value of Galactic reddening as $\ebv_{\rm MW}$ = $0.0278\pm0.0002$ mag. The reddening due to host
 galaxy $\ebv_{\rm host}$ was estimated using two methods viz. the blackbody approximations to 
 the 4d fluxes and the narrow blended \Nai\, doublet absorption lines in the spectra.  

\begin{figure}
 \centering
 \includegraphics[width=8.4cm]{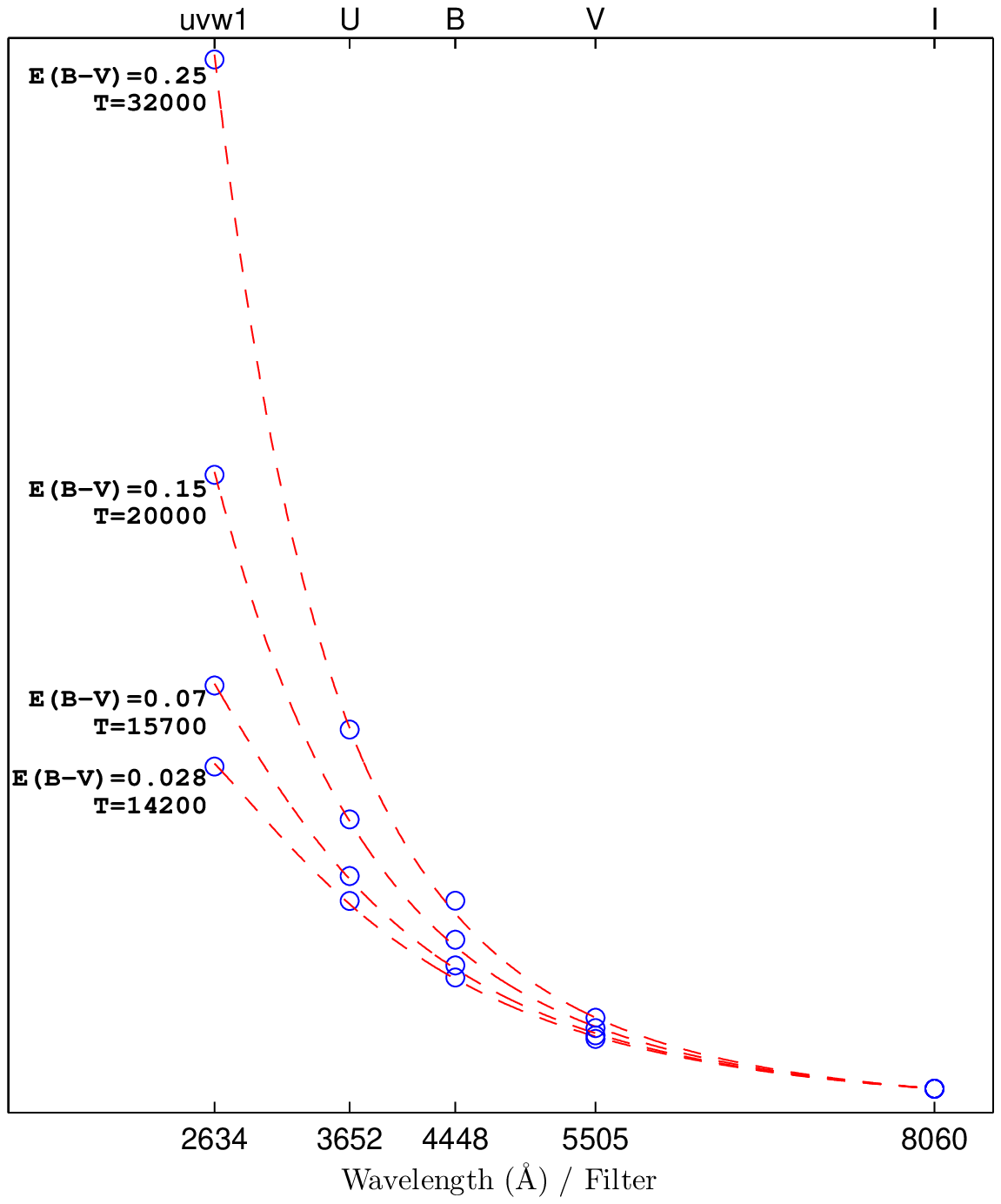}%
 \caption{The spectral energy distribution of \sn\, at 4d is compared with blackbody function. The 
          open circles denote observed fluxes in $uvw1$, $U$, $B$, $V$ and $I$ bands corrected
          for respective reddening values while the dotted line is the best-fit models for different 
          temperature. The fluxes are normalized relative to I-band flux.}
 \label{fig:ebv.sed}
 \end{figure}

 The observed spectral energy distribution (SED) of a few days old SNe can be approximated as 
 a blackbody and hence, we generated observed spectral fluxes between 
 0.26 \mum\, to 0.81 \mum\, using photometric data at 4d in $uvw1$ band ($\lambda_{c}$ = 2600 \AA) 
 of {\it Swift} Ultra-violet Optical Telescope (UVOT) taken from \citet{2013ApJ...764L..13B} 
 and in $UBVI$ bands from the present study. The 4d data for UVOT 
 filters $uvw2$ ($\lambda_{c}$ = 1928 \AA) and $uvm2$ ($\lambda_{c}$ = 2246 \AA) are 
 not used because of higher errors and also we did not rely on $R$ flux because of the 
 high contamination due to H$\alpha$ emission. In Fig.~\ref{fig:ebv.sed}, we have plotted the 
 dereddened observed fluxes for varying \ebv\, and here, the dereddening is done using 
 reddening law of \citet{1989ApJ...345..245C} for a total-to-selective extinction 
 ratio ($R_{\rm V}$) of 3.1. We also show the corresponding blackbody model fluxes with 
 best-fit temperatures; and corresponding to \ebv\, of 0.028 mag, we derive a temperature 
 of 14.2 kK. For \ebv\, = 0.25 mag, we obtain an unphysical high temperature of 32 kK. 
 The theoretical modelling of \citet{2006A&A...447..691D,2011ApJ...729...61B} indicate that 
 a value of temperature above 20kK is not expected for 4d old type IIP SNe and hence 
 we derive an upper limit for total \ebv\, of 0.15 mag.

\begin{figure}
\centering
\includegraphics[width=6cm]{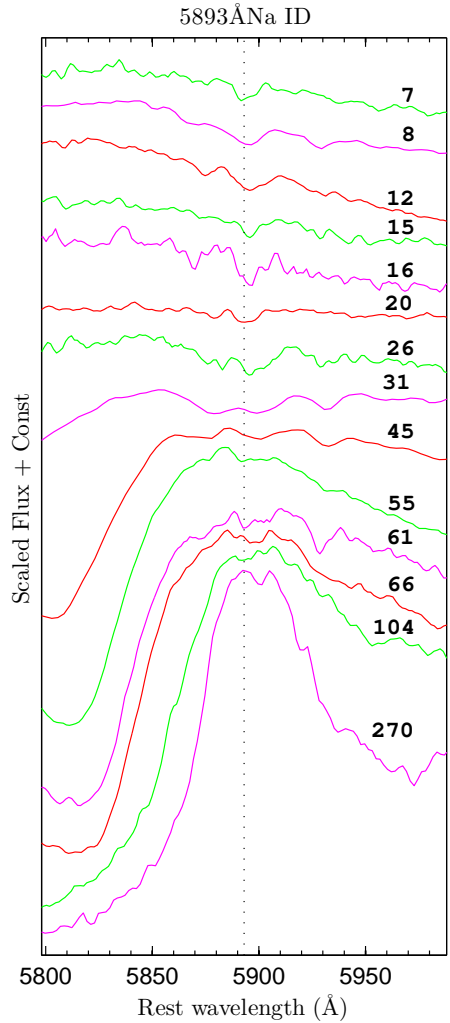}%
\caption{The \Nai\,D blended feature due to host galaxy at around 5893\AA\ is shown by dotted line.}
\label{fig:ebv.nad}
\end{figure}

 The equivalent width (EW) of \Nai\,D absorption feature is found to be correlated with the 
 reddening \ebv\, estimated from the tail of SN Ia color 
 curves \citep{1990A&A...237...79B,2003fthp.conf..200T}, though it may be a bad proxy in 
 certain cases, e.g. see \citet{2011MNRAS.415L..81P}. In the low-resolution spectra presented in 
 this work, it is not possible to resolve the individual component of \Nai\, doublet 
 (D$_{1}$ 5889.95\AA\, and D$_{2}$ 5895.92\AA) and hence we can expect to see the blended
 feature at the gf-weighted rest wavelength of 5893\AA. At all the 14 phases of spectra, a weak 
 impression of \Nai\,D due to the host is seen overlaid on the broad P Cygni profile due to 
 SN (see Fig.~\ref{fig:ebv.nad}), 
 whereas, the feature due to Milky Way is comparatively weak and it is visible only for 
 spectra with high SNR. In Table~\ref{tab:ew}, we list the EW of \Nai\,D due 
 to host galaxy only and no attempt is made to estimate EW due to Milky Way. Due to poor SNR 
 ($<50$; Table~\ref{tab:speclog}), the 16d, 104d and 270d spectra were not included 
 in the EW determination. The error in EW is calculated using the relation given 
 by \cite{2006AN....327..862V} for weak-line limit. The weighted mean of EW derived from 11 individual 
 measurements is $0.394\pm0.067$\AA. Employing empirical relation from \cite{2012MNRAS.426.1465P}, i.e.
 log$_{\rm 10}\ebv_{\rm host}=1.17 \times {\rm EW}-1.85\pm0.08$ (where EW in \AA), a value 
 of $0.041\pm0.011$ mag is obtained for reddening due to host galaxy and the error quoted here includes 
 both that in EW and that in the empirical relation. This value is consistent with 
 the $\ebv_{\rm host}=0.055\pm0.014$ mag which is derived by \cite{2012ApJ...756..131V} using 
 high-resolution Echelle spectra of \sn\, obtained from 10-m Keck telescope at 25d. We derive a weighted 
 mean of the two above measurements, i.e. $0.046\pm0.008$ mag for the host galaxy and by 
 adding the Galactic reddening, a total $\ebv_{\rm tot}=0.074\pm0.008$ mag is derived for {\sn} and 
 is adopted throughout this paper. This 
 corresponds to a visual extinction $A_V=0.23\pm0.03$ mag, assuming line-of-sight ratio 
 of total-to-selective extinction $R_{\rm V}=3.1$.

\begin{table}
  \caption{Equivalent widths of \Nai\,D absorption feature due to host galaxy. } 
  \label{tab:ew}

  \begin{tabular}{c c c} \hline UT Date&Phase$^{a}$&EW\\
     (yyyy-mm-dd)&(day)&\AA \\ \hline
    2012-03-22 &   7 &   0.415$\pm$0.204     \\
    2012-03-23 &   8 &   0.410$\pm$0.185     \\
    2012-03-27 &  12 &   0.400$\pm$0.233     \\
    2012-03-30 &  15 &   0.374$\pm$0.361     \\
    2012-04-04 &  20 &   0.399$\pm$0.172     \\
    2012-04-10 &  26 &   0.370$\pm$0.226     \\
    2012-04-15 &  31 &   0.364$\pm$0.189     \\
    2012-04-29 &  45 &   0.413$\pm$0.276     \\
    2012-05-09 &  55 &   0.363$\pm$0.237     \\
    2012-05-15 &  61 &   0.419$\pm$0.233     \\
    2012-05-20 &  66 &   0.400$\pm$0.272     \\ \hline 
    Weighted mean &  &   0.394$\pm$0.067     \\ \hline 
                                    
  \end{tabular}
  \newline
  $^{a}$ With reference to the time of explosion JD 24546002.59\\ 
 
\end{table}

\section{Optical light-curve} \label{sec:lc} 

 \subsection{Apparent magnitude light-curves} \label{sec:lc.app}

\begin{figure*}
\centering
\includegraphics[width=15.5cm]{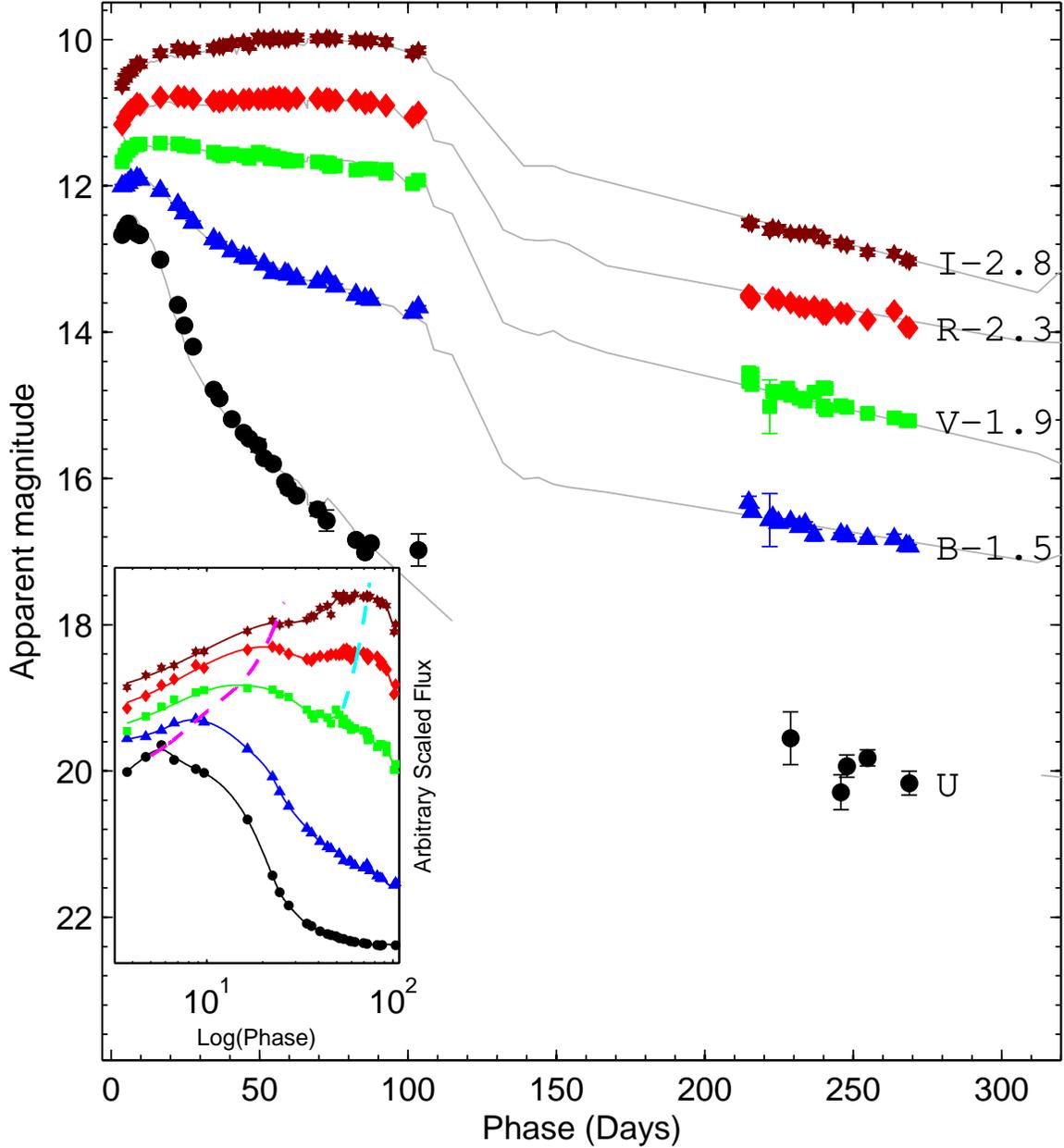}
\caption{The photometric light-curve in Johnson-Cousins $UBVRI$ system. The light-curves are shifted 
  for clarity, while for SN 1999em (gray solid lines), it is scaled in magnitude to match with \sn.
  The evolution of early-time time light-curve is shown in inset, wherein, the primary peaks in $U$, $B$, $V$,
  $R$, and $I$ at 6d, 9d, 14d, 20d and 23d respectively; and the secondary peak in $V$, $R$ and $I$
  at 52d, 56d, and 71d respectively are clearly visible. Primary and secondary peaks of the light-curve are connected by pair of dashed lines (magenta and cyan respectively).}
\label{fig:lc.app}
\end{figure*}

 The optical light-curve of {\sn} in $UBVRI$ filters is shown in Fig.~\ref{fig:lc.app}. The 
 photometric measurements are made at 54 phases during 4d to 269d. The SDSS $griz$ magnitudes
 are converted to $BVRI$ using empirical relations given by \citet{2006A&A...460..339J} and 
 are over-plotted For comparison the light-curves of archetypal type IIP 
 SN 1999em \citep{2002PASP..114...35L} is also shown. The early light-curve shows sharp 
 initial rise of brightness in all the optical bands and then declining slowly into 
 plateau-phase followed by a sharp fall at around 110d to the nebular-phase. On the contrary,
 the observations in UVOT bands do not show any initial rise in the light-curve observed since 
 as early as 3d \citep{2013ApJ...764L..13B}. The peak in early-time light-curve 
 occurs at about 8d, 11d, 15d, 22d, 24d respectively for $UBVRI$ bands, followed by a 
 continuous decline in $UB$ and a short decline then a continuous rise in $VRI$ peaking at 52d, 56d and 71d 
 respectively (see inset in Fig.~\ref{fig:lc.app}).  
 The early-time light-curve of \sn\, is almost similar to other nearby ($\le$ 11 Mpc), well-studied 
 normal type IIP SNe, e.g in SN 1999em - $UBV$ peaked at 6d, 8d and 
 10d \citep{2002PASP..114...35L}; in SN 2004et, $UBVRI$ peaked at 9d, 10d, 16d, 21d 
 and 25d \citep{2006MNRAS.372.1315S}; in SN 1999gi, $BV$ peaked at 8d and 
 12d \citep{2002AJ....124.2490L}. Appearance of initial peaks in $UBVRI$ seems to be a generic
 feature of early-time light-curves of IIP SNe and in absence of early-time data, it can provide
 a good handle on estimating the time of explosion with accuracy of a few days, e.g. at V-band, 
 the initial peak occurs at $13\pm3$ d for well studied IIP SNe. Unlike other SNe, 
 the densely-sampled light-curve of \sn\, offers the opportunity to see, for the first 
 time, the minima near 42d in $V$, 39d in $R$, 31d in $I$ band and then a slow rise to a plateau
 at about 51d, 59d and 73d respectively. The change in flux from initial early-time rise to the minima
 is about 16\% ($\sim$0.18 mag) in $ V $, 6\% ($\sim$0.07 mag) in $ R $, 2\% ($\sim$0.02 mag) in $ I $. 
 We note that 
 these values are larger than the typical photometric errors at these epochs. This observed minima 
 around which the SN cools down to hydrogen recombination temperature $\sim$ 6000 K, most likely marks 
 the tail-end of the flux from the adiabatic cooling phase of the shock-breakout and the dominance of 
 the flux from the hydrogen recombination phase in the supernova 
 envelope \citep{2009ApJ...703.2205K,2010APh....33...19C, 2011ApJ...736...76R}. 
 However, a dense coverage of this early time data for more number of events would be required 
 to confirm the nature and exact cause of this re-brightening in the light-curves.  

 As the SN goes behind the sun, we have a data gap between 105d to 214d, however, a prolonged
 plateau phase of about 100d is apparently seen. The decline rates up to 104d after maxima in $UBV$ is 
 5.60, 1.74, 0.55 mag/100d respectively, which is similar to the values observed for SN 1999em 
 (see Fig.~\ref{fig:lc.app}) and to SN199gi \citep{2002AJ....124.2490L}. The rate of decline for 
 SN 2004et is a bit higher, e.g. at B-band it is 2.2 mag/100d. During plateau-phase, R-band 
 light-curve shows almost no change in brightness, whereas the I-band shows slow 
 increase in brightness until mid-plateau of $\sim60$ days, then it remains almost constant 
 until the end of plateau. The decline rate (mag/100d) in light-curve of the nebular phase 
 between 215d to 269d is estimated as 1.24, 0.88, 0.88, 0.81, and 0.95 respectively
 for $UBVRI$.

 \subsection{Absolute magnitude and color evolution} \label{sec:lc.abs}

\begin{figure}
\centering
\includegraphics[width=8.5cm]{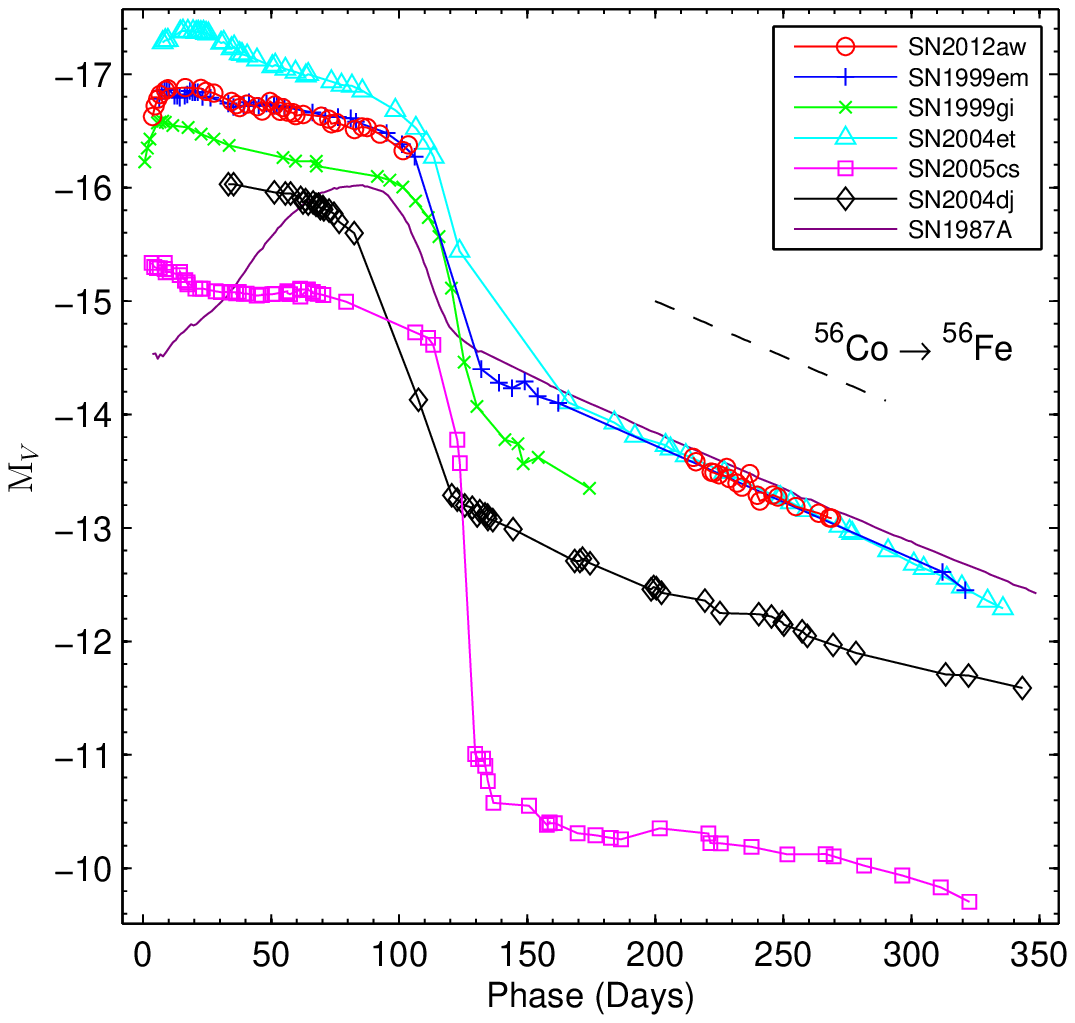}%
\caption{Comparison of M$_{V}$ light curve of {\sn} with other type IIP SNe. The exponential decline 
         of the radioactive decay law is shown with dashed lines. The time of explosion in JD-2400000, 
         distance in Mpc, reddening \ebv\, in mag  and 
  the reference for apparent V-band magnitude, respectively, are : 
  SN 1999em -- 51475.6, 11.7, 0.10; \citet{2002PASP..114...35L,2003MNRAS.338..939E};
  SN 2004et -- 53270.5, 5.4, 0.41; \citet{2006MNRAS.372.1315S}; 
  SN 2005cs -- 53549.0, 7.8, 0.11; \citet{2009MNRAS.394.2266P}; 
  SN 2004dj -- 53187.0, 3.5, 0.07; \citet{2008PZ.....28....8T}; 
  SN 1987A  -- 46849.8, 0.05, 0.16;  \citet{1990AJ.....99.1146H}; 
  SN 1999gi -- 51522.3, 13.0, 0.21; \citet{2002AJ....124.2490L}.}
\label{fig:lc.abs}
\end{figure}

 After correcting for distance and extinction (\S\ref{sec:ext}; Table~\ref{tab:host}), the absolute 
 magnitude $V$-band light-curve is shown in Fig.~\ref{fig:lc.abs} and it is compared with other well-studied 
 SNe of normal type IIP 1999em, 2004et, 1999gi, 2004dj; subluminous type IIP 2005cs and peculiar type II
 1987A. The $V$-band light-curves of SNe from literature are also corrected for distance and extinction and 
 the time of explosion for all (except SN 2004dj) of them is known with an accuracy of a day. The comparison 
 shows striking 
 similarity with SN 1999em in both shape and flux. If we ignore the effect of distance, then the 
 duration of plateau, the nebular phase luminosity and the overall shape of light-curve are similar for
 SNe 1999em, 1999gi, and 2012aw; in contrast to that seen for SNe 2005cs and 2004et.
 For \sn\, the plateau-to-nebular phase transition occurs at $\sim$ 117d and the mid-plateau 
 M$^{p}_{V}$ is $-16.66$ mag and hence it belongs to a normal type IIP 
 events \citep{1994A&A...282..731P}, in contrast to the subluminous IIP like 2005cs 
 with M$^{p}_{V}$ $\sim$ $-15$ mag \citep{2009MNRAS.394.2266P}.
 The peak absolute magnitude of \sn\, is equal to that of SN 1999em, about 0.31 mag brighter than
 SN 1999gi and about 0.49 mag fainter than SN 2004et. The light-curve  evolution in the nebular phase
 follows the decay rate 0.92 mag/100d. The photometric evolution of type IIP SNe in nebular phase 
 is powered by the radioactive decay of \cobalt\, into \iron\, and the expected decay rate 
 is 0.98 mag (100d)$^{-1}$. The values for \sn\, is consistent with this.

\begin{figure}
 \centering
 \includegraphics[width=8.5cm]{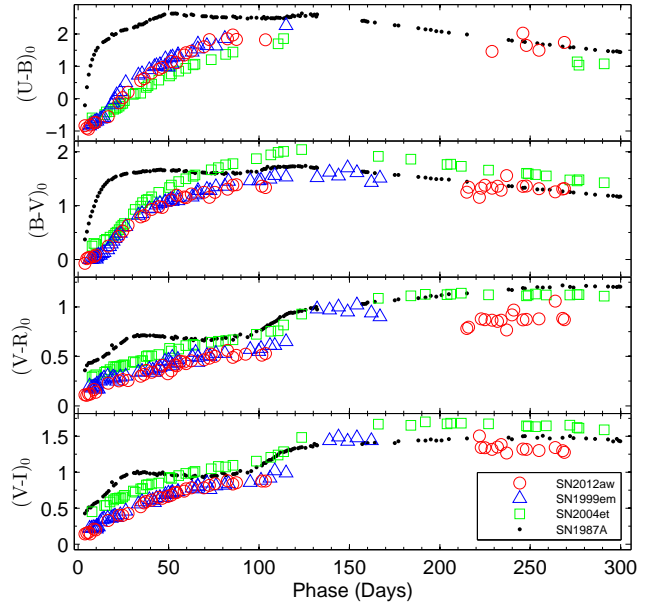}
 \caption{The evolution of intrinsic colors of {\sn} is compared with other well-studied normal 
          type IIP SNe 1999em, 1999gi and 2004et. The reference for the data is same as in 
          Fig.~\ref{fig:lc.abs}.}
 \label{fig:cc.abs}
 \end{figure}

 The evolution of broad-band colors provides important clues for the expansion and cooling behavior of 
 the supernova envelope. In Fig.~\ref{fig:cc.abs}, the evolution of intrinsic colors $U-B$, $B-V$, $V-R$ 
 and $V-I$ are shown. The $U-B$ color evolves very rapidly in early phases up to $\sim$ 50d, primarily 
 due to high temperature and rapid cooling; and it becomes redder from $-0.94$ to 1.11 mag; though it evolves 
 slowly thereafter and reaches 1.96 mag by 90d. By 260d, it becomes 
 blue again by 0.6 mag in about 150d. The $B-V$ also evolves similar to $U-B$, and in nebular phases both
 colors follow the evolution completely similar to SN 1987A. The evolution in all the colors shows 
 striking resemblance to that of SN 1999em. The $V-I$, $V-R$ and $B-V$ colors of \sn\, are significantly 
 ($\sim$ 0.2-0.8 mag) bluer than that of SN 2004et at all phases, while the $U-B$ color is observed to be 
 redder.

 \begin{figure}
 \centering
 \includegraphics[width=8.4cm]{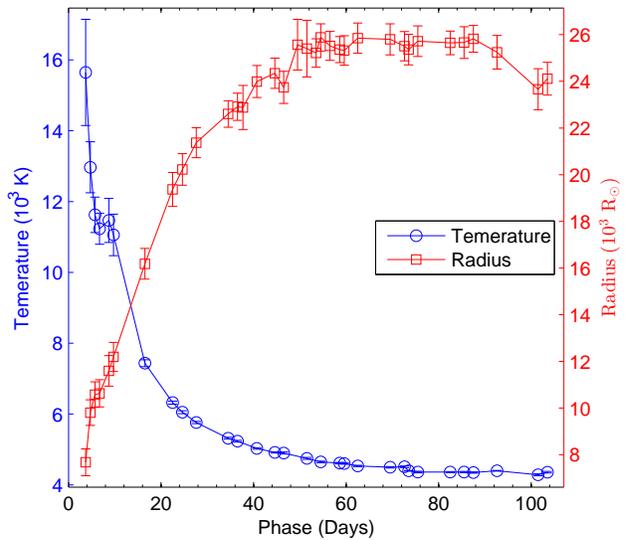}%
 \caption{The temperature and radius of \sn\ as derived from blackbody fits to 
          the observed fluxes in the ultraviolet-optical range 0.17-0.88 $\mum$.}
 \label{fig:lc.rad}
 \end{figure}

 In order to have an idea on the temporal evolution of temperature, we fitted blackbody on the 
 observed fluxes in $uvw1$ band \citep[data taken from][]{2013ApJ...764L..13B} and in $UBVRI$ 
 bands covering the wavelength region 0.26-0.81 $\mum$. The $uvm2$ and $uvw2$ bands are not 
 used as these have large uncertainties in magnitudes and also they result in unrealistic
 blackbody fits at early phases. The fluxes were corrected for interstellar 
 extinction and we also applied dilution factors as per the prescription by \citet{2006A&A...447..691D}. 
 The fitted values of temperature and the radius are plotted in Fig.~\ref{fig:lc.rad}. The blackbody 
 temperature ($T_{\rm bb}$) thus derived can be approximated as a photospheric temperature. 
 During the phases 4-10d, the ejecta temperature drops from 16 kK to 11 kK and by 20d it drops down 
 to about 6500 K, with a very slow decline 
 thereafter to 4300 K by 104d. This indicates that the plateau seen in the absolute magnitude light-curve 
 between $\sim$ 20d to 104d is mainly sustained by the recombination of hydrogen. The sharp change 
 in slope of $T_{\rm bb}$ around 20-30d is consistent with the similar trend seen in the color-curve 
 evolution of $U-B$ and $B-V$.


 \subsection{Bolometric light-curve} \label{sec:lc.bol}

 \begin{figure}
 \centering
 \includegraphics[width=8.5cm]{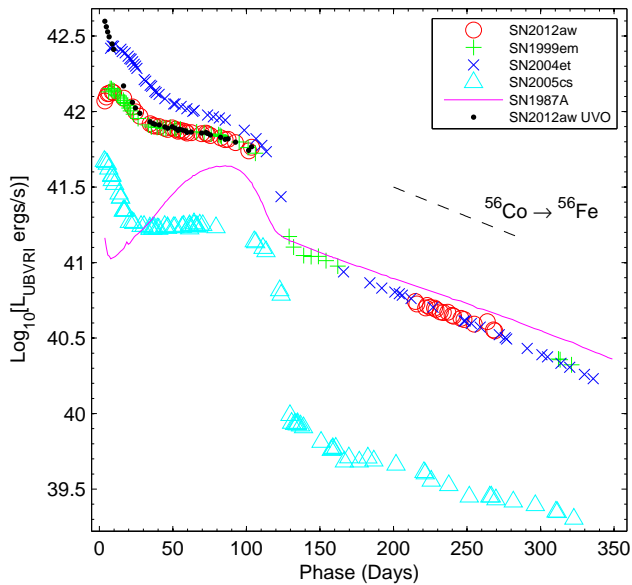}%
 \caption{The $UBVRI$ bolometric light-curve of {\sn} is compared with other well studied supernovae. 
          The adopted distances, reddening and time of explosion are same as in Fig.~\ref{fig:lc.abs}. The
          exponential decline of the radioactive decay law is shown with dashed lines.}
 \label{fig:lc.bol}
 \end{figure}


 The bolometric light-curves provide constraint on the amount of radioactive \nickel\, synthesized during 
 explosion as well as on the energy of supernova explosion. At early phases ($\le$ 30d), when supernova 
 ejecta is hotter, the major
 contribution ($\sim 80\%$) to the bolometric light comes from the ultraviolet and $UB$ bands, while at 
 later phases the flux contribution from $RI$ and near-infrared band dominates \citep{2007MNRAS.381..280M}.   
 Here, we determine pseudo-bolometric light-curve by integrating a spline fitted on the $UBVRI$ 
 fluxes derived at their respective effective wavelengths using zero-points from \cite{1995PASP..107..945F}. 
 Fluxes are integrated over wavelength range from 3335 \AA\, to 8750 \AA; the lower and upper bounds are 
 extended to the HWHM (half width half maximum) of $U$ and $I$ bands. Wherever, the observations in a 
 bandpass were missing, the 
 magnitudes were obtained by interpolating the light-curves using low-order cubic-spline. In order to remove 
 the effect of overlapping wavelengths in the $UBVRI$ passbands on the 
 determination of SED from observed fluxes, we equated the known splines convolved with the filter 
 response with that of the observed fluxes. As a guess, we provide initial SED as the spline fitted 
 on observed fluxes and iteratively we construct the true spline SED (whose filter-convolved fluxes 
 matches with observed fluxes). In Fig.~\ref{fig:lc.bol}, we
 plot the bolometric light-curve of \sn\, and in the same figure we also show the bolometric light
 derived using UVOT fluxes in $ubw2$ and $uvw1$ bands taken from \citet{2013ApJ...764L..13B}. 
 The $uvm2$ flux is not used due to less number of measurements and large uncertainties in magnitudes. 
 To ensure proper comparison, we have applied same scheme to compute bolometric light-curve in $UBVRI$ bands
 for other well-studied type II SNe 1987A \citep{1990AJ.....99.1146H}, 
 1999em \citep{2002PASP..114...35L}, 2004et \citep{2006MNRAS.372.1315S} and 2005cs \citep{2009MNRAS.394.2266P}. 
 Similar to color-curve, the evolution of bolometric luminosity is also similar to SN 1999em. 
 The bolometric luminosity declines rapidly by 0.5 dex in first 30d of evolution, and 
 then it slowly declines by 0.2 dex by 104d.

\begin{figure}
\centering
\includegraphics[width=8.5cm]{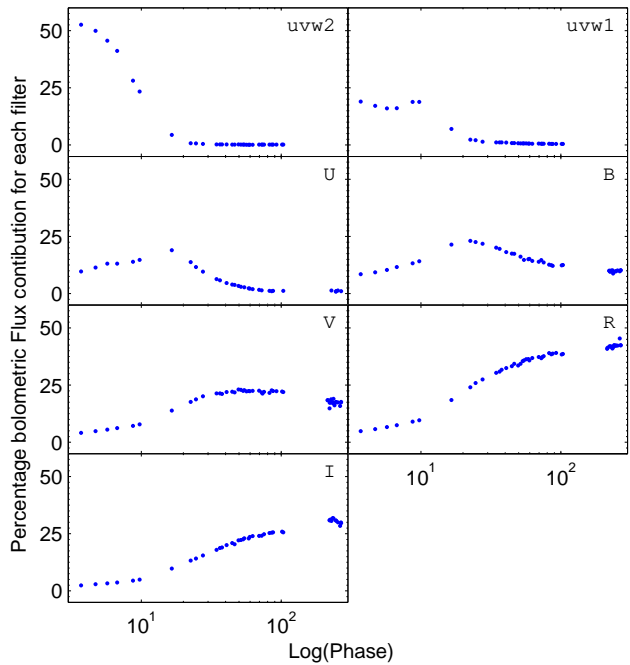}%
\caption{The percentage flux contribution in different passbands in the UVOT $uvw2$, $uvw1$ and optical $UBVRI$.
         Fluxes are corrected for extinction.}
\label{fig:lc.per}
\end{figure}

 In Fig.~\ref{fig:lc.per}, we plot the percentage flux contribution in reference to the total bolometric
 light in the ultraviolet-optical region (0.17 to 0.88 $\mum$) from the different passbands. The 
 contribution due to UVOT bands and the optical $UB$, show a sudden change at around 20d. In uvw2, 
 the contribution falls from approximately 50\% to 0\%; in uvw1, it drops to 0\% from a level of 25\%; 
 while in U and B, it rises from 10\% at 4d to 22\% at about 20d and falls slowly thereafter to a few 
 percent level in plateau and nebular phases. In $V$, $R$ and $I$ bands, the trend in flux contribution
 behaves in a similar fashion and it slowly rises from a few \% at 7d to above 25\% in the plateau
 phases and beyond. The bolometric luminosity at 25d and beyond is mainly contributed by the $VRI$
 bands.

 \subsection {Mass of nickel} \label{sec:lc.nick}


 In type IIP SNe, the radioactive \nickel\ is synthesized by explosive burning of Si and O during shock 
 breakout phase of the explosion \citep{1980ApJ...237..541A} and hence the nebular phase 
 light-curve is mainly powered by radioactive decay of \nickel\, to \cobalt\ with half-life time
 of 6.1d and \cobalt\ to \iron\, with $ e $-folding time of 111.26d emitting energetic $\gamma$-rays and 
 positrons. The nebular phase luminosity is assumed to be directly proportional to mass of \nickel. 
 For a nearby type II-pec SN 1987A, the mass of \nickel\, ($M_{\rm Ni}$) has been determined fairly accurate to 
 be $0.075\pm0.005$ \msun\ \citep{1996snih.book.....A}, and hence by assuming similar $\gamma$-ray energy deposition and 
 by comparing the bolometric luminosity of SN 1987A at comparable epochs, we can estimate $M_{\rm Ni}$ 
 of \sn\,. We note that for comparison, we use $UBVRI$ bolometric luminosity of 1987A computed in 
 this work (\S\ref{sec:lc.bol}) instead of already available accurately calculated UV-optical-IR 
 luminosity. For {\sn}, the $UBVRI$ bolometric luminosity at 240d is estimated using a 
 linear fit over 18 epochs between 210d to 270d, which 
 is $4.53\pm 0.11\times 10^{40}$\ergs. Similarly for 1987A, it is 
 estimated to be $5.82\pm 0.06\times 10^{40}$\ergs. The ratio of {\sn} 
 to 1987A is found to be $0.778\pm0.021$ and hence we calculate the value of $M_{\rm Ni}$
 for {\sn} to be $0.058\pm0.002$ \msun\footnote{ 
 Using the same method and adopting the distance and extinction as given in Fig.~\ref{fig:lc.abs}, we 
 estimate $M_{\rm Ni}$ for comparison SNe 2005cs, 1999em and 2004et as $0.004\pm0.001$ \msun, 
 $0.053\pm0.001$ \msun, and $0.058\pm0.001$ \msun\ respectively.}
 
 The value of $M_{\rm Ni}$ can be independently estimated using the tail luminosity ($L_{\rm t}$) as described 
 by \cite{2003ApJ...582..905H}, assuming the $\gamma$-rays emitted during radioactive decay of \cobalt\ 
 makes the ejecta thermalized viz.
 \begin{eqnarray*} 
  M_{\rm Ni} = 7.866\times10^{-44} \times L_{t} \exp\left[ \frac{(t_{t}-t_{0})/(1+z)-6.1}{111.26}\right]\msun
 \end{eqnarray*}
 where $t_{0}$ is the explosion time, 6.1d is the half-life of \nickel\ and 111.26d is the e-folding time of the 
 \cobalt\ decay emitting energy in the form of $\gamma$-rays. We compute $L_{t}$ at 11 epochs 
 between 215 to 265 days using the observed $V$ magnitude, a bolometric correction 
 of $0.26 \pm 0.06$ mag during nebular phase \citep{2003ApJ...582..905H}, and the adopted 
 reddening and distance (see Table~\ref{tab:host}). The weighted mean of 
 $L_{\rm t}$ is  $8.97\pm0.19\times10^{40}\,$\ergs corresponding to mean phase of 240.35d,
 and this results in a value of $M_{\rm Ni} =0.057\pm0.012$\msun. 
 Taking weighted average of the values derived above using two photometric methods, we adopt value 
 of $M_{\rm Ni}$ as $0.058\pm0.002$\msun\ for \sn. 

 We note that the value of $M_{\rm Ni}$ estimated for type IIP SNe using bolometric 
 luminosity of nebular phase depends considerably on the adopted distance and extinction.
 For example, the value of $M_{\rm Ni}$ reported in the 
 literature \citep[][see their Table 1, hereafter TV12]{2012MNRAS.419.2783T} 
 for SNe 2005cs, 1999em and 2004et varies from 0.003 to 0.008\msun, from 0.022 to 0.036\msun, 
 and from 0.056 to 0.068\msun\ respectively. However, in view of the empirical correlation found between 
 mass of \nickel\ and mid-plateau photospheric velocity by \citet{2003ApJ...582..905H} using a 
 large sample of type IIP SNe, the observed object-to-object variation in mass of \nickel\ for the 
 above three cases appears to be realistic as the mid-plateau ($\sim$50d post explosion) photospheric velocity 
 of SNe 2005cs, 1999em and 2004et estimated using \synow\ modelling of spectra is found to be 
 1200\kms, 3400\kms\ and 3750 \kms\ respectively (TV12). Considering the uncertainty in the 
 adopted distance and extinction for \sn\, we anticipate that it produced the amount of \nickel\ equal to or 
 a bit less than that for SN 2004et and this is further corroborated by the fact that the mid-plateau 
 photospheric velocity of \sn\ is found to be higher than that of SN 1999em but similar to SN 2004et at 
 comparable epochs (\S\ref{sec:sp.vel}).


\section{Optical spectra} \label{sec:sp}


\subsection{Key spectral features} \label{sec:sp.key}
%
%

\begin{figure*}
\centering
\includegraphics[width=17cm]{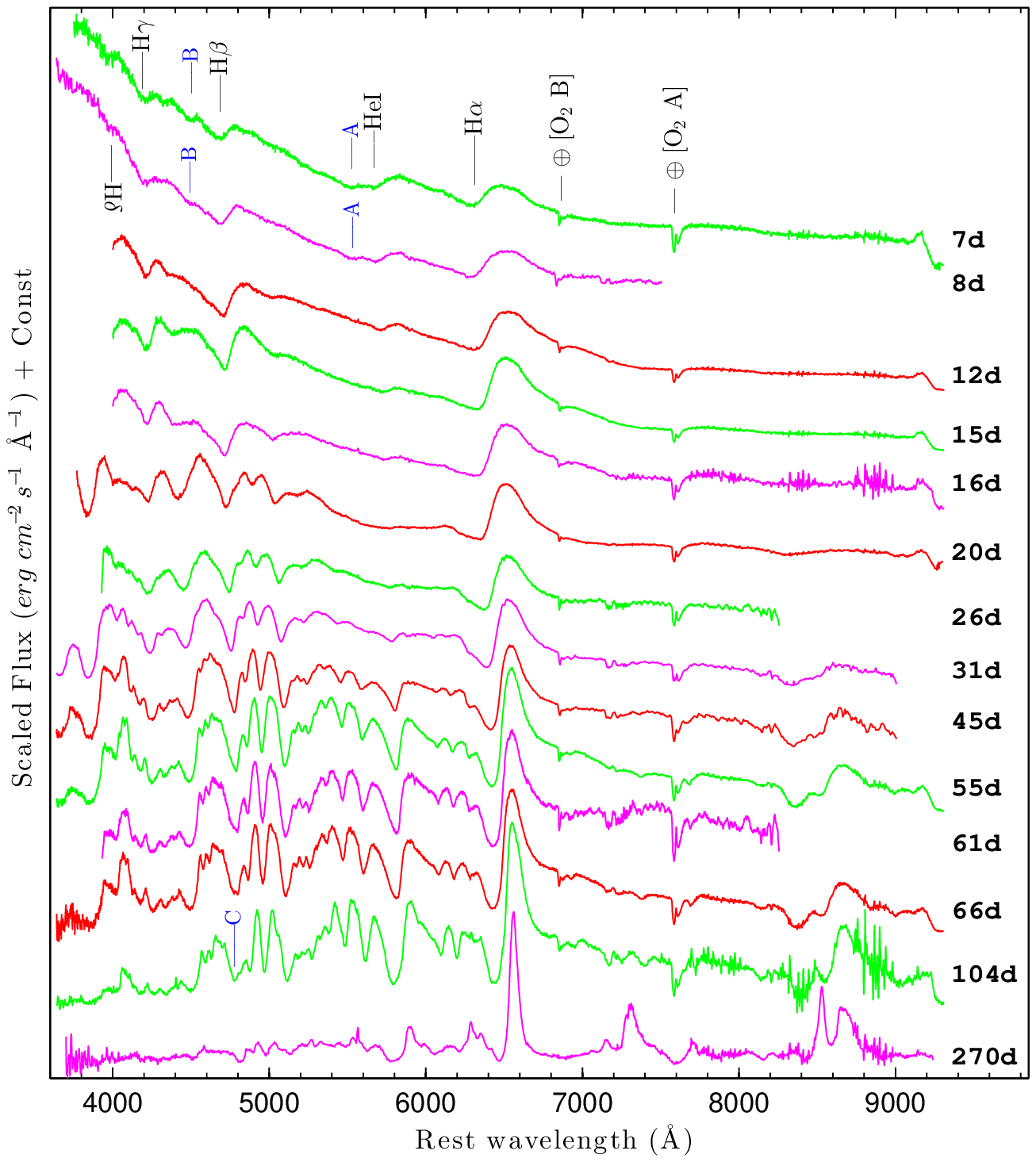}
\caption{The Doppler corrected spectra of \sn\ are shown for 14 phases during 7d to 270d. The prominent P-Cygni 
         profiles of hydrogen (\ha, \hb, \hg, \hd) and helium (\Hei\ 5876\AA) lines are marked. The peculiar 
         absorption features marked with A, B and C are discussed in text. The telluric absorption features 
         are marked with \earth\, symbol. The portions of spectrum at extreme blue and red end have poor 
         signal-to-noise ratio.}
\label{fig:sp.all}
\end{figure*}

 The spectroscopic evolution of \sn\ is presented in Fig.~\ref{fig:sp.all}, in which a preliminary 
 identifications of spectral features is done as per the previously published lines for type IIP events
 by \citet{2002PASP..114...35L} and the absorption component of some prominent lines are marked. All
 the spectra are corrected for recession velocity of the host galaxy (\S\ref{sec:obs.spec}). 
 The early phase (7d and 8d) spectra can be distinguished with featureless 
 blue continuum having broad P-Cygni profiles of hydrogen \Hi\ (\ha\ 6562.85\AA, \hb\ 4861.36\AA,
 \hg\ 4340.49\AA, \hd\ 4101.77\AA) and helium \Hei\ 5876\AA. The \Hei\ line disappears 
 by $\sim$ 16d as the continuum becomes redder with time corresponding to a sudden drop 
 in $T_{\rm bb}$ by a factor of two to 7 kK since early phase. The appearance of \Hei\ line since
 early phases and its disappearance exactly at $\sim$ 16d is also seen 
 in SN 1999em \citep{2002PASP..114...35L}. The Balmer lines of \Hi\ are seen at 
 all the phases until 270d. The FWHM of the emission component of \ha\ decreases from 
 $\sim$ 17000\kms\ at 7d to about 2500\kms\ at 270d, indicating a decrease in temperature and 
 opacity of the \Hi\ line-emitting regions.

 In addition to the regular features of \Hi\ and \Hei, the early (7d and 8d) spectra have a couple of 
 peculiar absorption features. Two weak absorption features are present at $\sim$~4300\AA\
 and $\sim$~4850\AA\ while two relatively strong absorption features are seen 
 near $\sim$~5500\AA\ (marked with A) and near 4500\AA\ (marked with B).
 Similar features have also been identified in the early spectrum of 
 type IIP SNe 1999em \citep{2002PASP..114...35L}, 1999gi \citep{2002AJ....124.2490L},
 and 2007od \citep{2011MNRAS.417..261I}. The origin of these features is not completely 
 understood and in most of the cases they are explained as high-velocity component of 
 \Hi\ and \Hei\ lines, however, the presence of \Nii\ lines 4623\AA\, 5029\AA\ and 5679\AA\ 
 have also been explored and these lines cannot be completely ruled 
 out \citep{2000ApJ...545..444B,2012MNRAS.422.1178I}.   

\begin{figure}
 \centering
 \includegraphics[width=8.4cm]{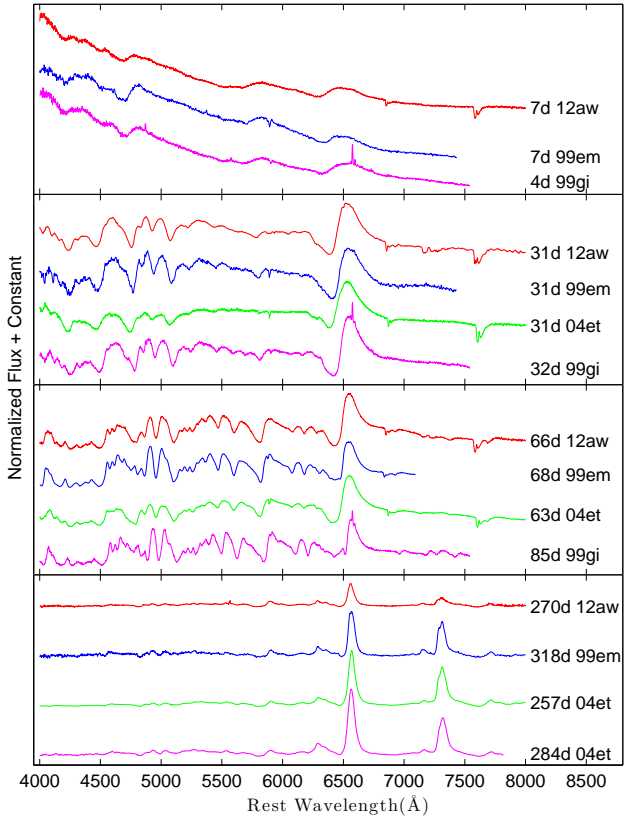}%
 \caption{Comparison of early (7d),  plateau (31d, 66d) and nebular (270d) phase spectra of {\sn} with 
          other well-studied type IIP SNe 1999em \citep{2002PASP..114...35L}, 1999gi \citep{2002AJ....124.2490L}, 
          2004et \citep{2006MNRAS.372.1315S,2010MNRAS.404..981M}. Observed fluxes of all 
          the SNe are corrected for extinction and redshift (adopted values same as in Fig.~\ref{fig:lc.abs}).}
 \label{fig:sp.lit}
 \end{figure}

 In \sn, the feature A at 7d is at $\sim17665$ \kms\ with reference to \Hei\ rest wavelength, thus making 
 it $\sim7198$ \kms\ higher over the existing \Hei\ absorption feature velocity. This component 
 decreases to $\sim16643$ \kms\ at 8d  making an offset by $\sim6586$ \kms\  higher 
 than existing \Hei\ feature. Another absorption feature B is also located at a velocity position 
 of $\sim21785$ \kms\ on 7d in reference to rest \hb\, making it higher 
 by $\sim11293$ \kms\ over the existing \hb\ absorption trough. This feature decreases 
 to $\sim21477$ \kms\ on 8d. Both of these features tend to decrease in velocity with 
 time, however, none of these features could be detected on 12d and after.
 The HV absorption features are not seen in \ha. In Fig.~\ref{fig:sp.lit}, we compare the early-phase
 spectra with other SNe and it can be seen that such HV features indicating existence of very
 high-velocity line-forming material have also been observed for SN 1999gi \citep{2002AJ....124.2490L} 
 and SN 1999em \citep{2000ApJ...545..444B,2002PASP..114...35L}. In the day 1 spectra of SN 1999gi,
 the HV component was more dominant and it appeared at $-30000$ \kms\ with no trace of normal absorption
 components. \cite{2012MNRAS.422.1178I} detect HV absorption to \hb\ and \hg\ lines only in a bright (Mv = -18) type IIP
 SN 2007od, and a detailed spectral analysis using \phoneix\ favored presence of HV feature. It appears 
 that such high-velocity features 
 in \hb\ and \Hei\ are ubiquitous in normal luminosity type IIP SNe at early phases ($\le$ 8d) though 
 its origin (i.e. whether it is caused by abundance and/or density enhancements in the layers of ejecta above
 photosphere) and exact geometry remains open questions to be addressed using detailed modelling.       
 We attempt to identify these peculiar absorption features using \synow\ modelling in \S\ref{sec:sp.syn}.

 The spectra at 12d, 15d and 16d represent transition from a hotter ($\sim$16kK) early phase to a 
 cooler ($\sim$6kK) plateau phase when photosphere begin to penetrate Fe-rich ejecta. These spectra mark 
 the emergence of permitted lines of singly ionized atoms of calcium, iron, scandium, barium, titanium and
 of neutral sodium atoms. The strong lines of \Baii\ 4554\AA\ blend and \Feii\ 5169\AA\ 
 begin to appear at 12d, and the weaker lines of \Feii\ at 4929\AA\ and 5018\AA\ are clearly seen at 20d.
 The spectra at 8 epochs from 20d to 104d represent plateau-phase corresponding to further slow cooling of 
 the supernova envelope and appearance of more number of metallic lines. The lines of \Nai\ D 5893\AA\ 
 and \Caii\ IR Triplet 8498, 8542, 8662\AA\ emerge at 20d and they are clearly visible by 31d.
 The blend of \Caii\ H 3934\AA\ \& K 3968\AA\ appear at the gf-weighted rest wavelength 3945\AA\ and it is 
 very strong at 20d; seen until 104d. A definite
 identifications of \Caii\ is seen in early spectrum of IIP SNe as well, e.g. at 12d 
 in SN 1999em \citep{2000ApJ...545..444B}. The \Feii\ 5535\AA\ blend, \Scii\ 5665\AA\ multiplet, 
 \Baii\ 6142\AA, and \Scii\ 6246\AA, are clearly seen at 45d.
 The above features are present in the spectra until 104d and their comparison with other 
 SNe (Fig.~\ref{fig:sp.lit}) indicate that the plateau phase spectral features are similar to
 normal type IIP events. 

\begin{figure}
\centering
\includegraphics[width=8.5cm]{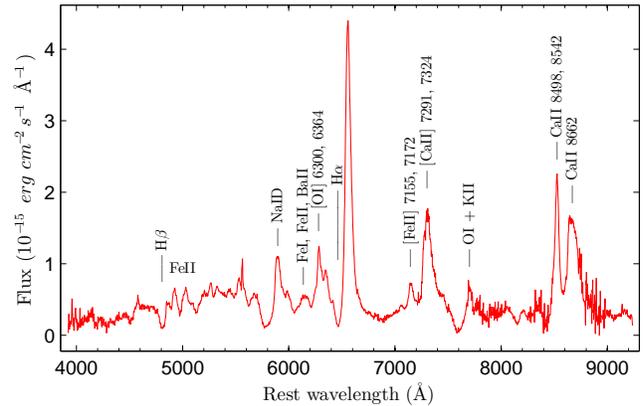}%
\caption{The nebular phase spectrum of \sn\ at 270d.}
\label{fig:sp.neb}
\end{figure}

 The only late time spectrum at 270d shows emission lines with no absorption components, which is a typical 
 characteristic feature of nebular phase spectra of IIP SNe. A comparison of 270d spectrum are made 
 with other SNe in Fig.~\ref{fig:sp.lit} and a preliminary identification of nebular lines are
 shown in Fig.~\ref{fig:sp.neb}. The forbidden emission lines of \Oia\ 6300, 6364\AA, 
  \Caiia\ 7291, 7324\AA\, and \Feiia\ 7155, 7172 \AA ; permitted emission lines 
 of \Hi, \Nai\ 5893\AA\ doublet and \Caii\ 8600\AA\ triplet become the dominant spectral features. 

 The presence and evolution of spectral features of \sn\ during early, plateau and nebular phases 
 show striking similarity with other well-studied normal type IIP SNe 1999em \citep{2002PASP..114...35L}, 
 1999gi \citep{2002AJ....124.2490L}, and 2004et \citep{2006MNRAS.372.1315S}. 
 However, in order to identify and interpret weak spectral features; to estimate the velocities 
 of photospheric/ejecta layers and to understand how the layers of line-forming regions evolve 
 with time and vary among object-to-object, we perform \synow\ modelling of the spectra in the next section.

\subsection{\synow\ modelling of spectra} \label{sec:sp.syn}

 We modeled the spectra of \sn\ with parameterized supernova spectrum 
 synthesis code \texttt{synow 2.3}\footnote{http://www.nhn.ou.edu/$\sim$parrent/synow.html} 
 \citep{1997ApJ...481L..89F,1999MNRAS.304...67F,2002ApJ...566.1005B}. In contrary to the 
 full non-local thermodynamic equilibrium (\nlte) model 
 codes \cmfgen\ \citep{2005ASPC..332..415D} and \phoneix\ \citep{2004ApJ...616L..91B}, 
 \synow\ assumes - simple \lte\ model atmospheres having a sharp photosphere emitting 
 a blackbody continuum; a spherically symmetric supernova expanding homologously;  
 the line formation is due to pure resonant scattering and radiative transfer is computed by employing 
 Sobolev approximation.
 We tried three different options for optical depth 
 profiles (viz., Gaussian, exponential and powerlaw), no significant 
 differences were noticed, however while matching absorption minimum the exponential 
 profile, $\tau\propto exp(-v/v_e)$, were $v_{e}$, a profile fitting parameter, e-folding
 velocity, was found to be most suitable and is adopted here.
 One important aspect of \synow\ model  is detachment of an ion (line-forming layer) from photosphere 
 which is achieved by setting the minimum velocity of the ion greater than 
 photospheric velocity resulting
  into flat-topped emission and blue-shifted 
 absorption counterpart. Strong \ha\ P-Cygni can be partially fitted in absorption minima with 
 detached \ion{H}{I} only, while the strong emission part remain unfitted with flat-topped crest.

\begin{figure*}
\centering
\includegraphics[width=\linewidth]{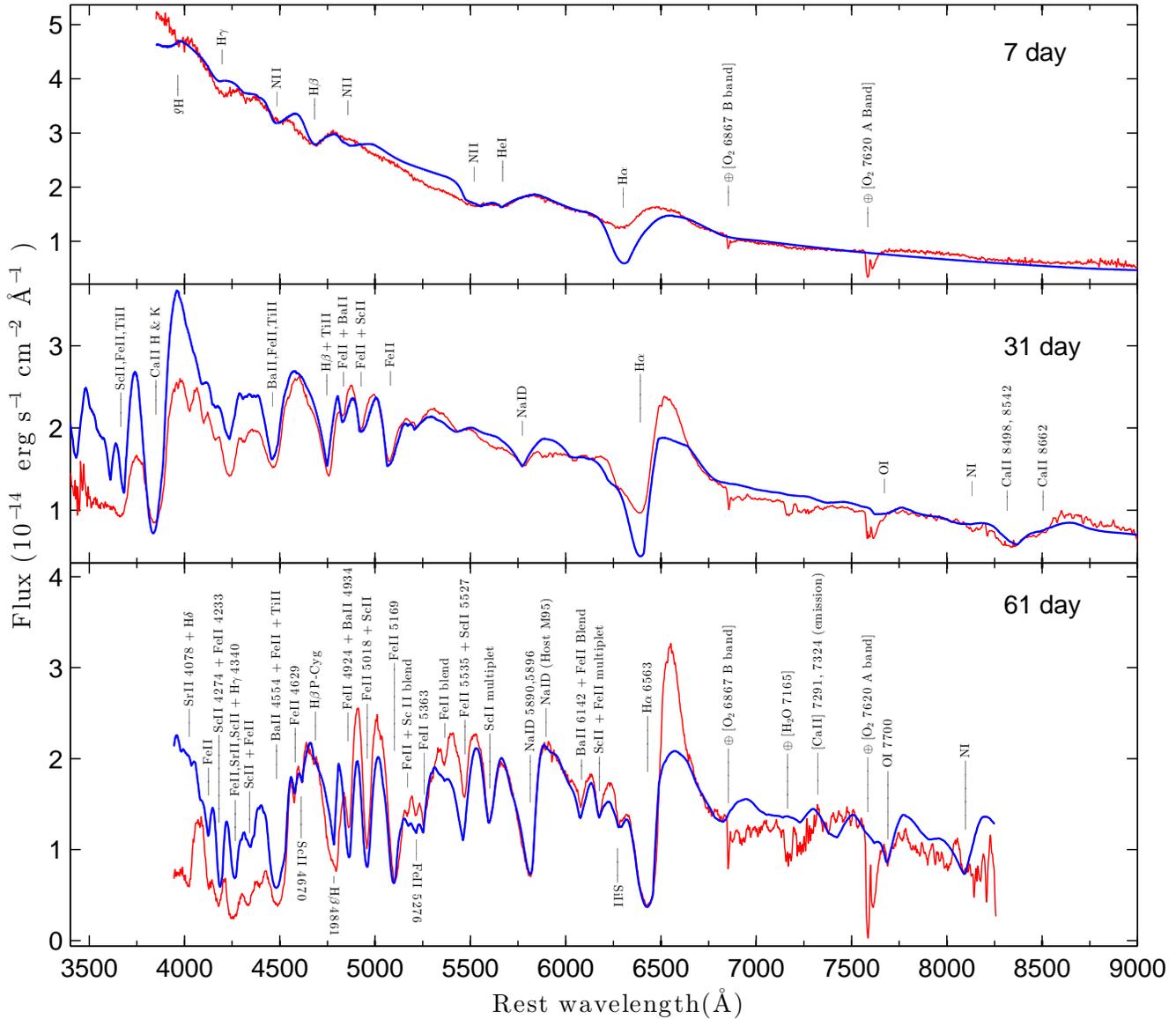}%
\caption{\synow\ modelling of 7d, 31d and 61d spectra of \sn. Model spectra are shown with thick solid line (blue),
         while the observed ones are in thin solid line (red). Observed fluxes are corrected for extinction.}
\label{fig:sp.synph}
\end{figure*}

 The observed spectra are Doppler-corrected and dereddened, before visually fitting the model 
 \synow\ spectra with the observed ones. We have modeled the spectra for all the 14 phases and 
 the best-fit values of $T_{bb}$ to match the continuum are given in Table~\ref{tab:synow}. 
 These \synow-derived temperatures are consistent with the ones derived photometrically (\S\ref{sec:lc.abs}). 
 In the early-phase spectra continuum is fitted properly as it agrees well with the \synow's 
 approximation of LTE atmosphere, whereas at later phases, the continuum is quite deviant, 
 though we are able to match the spectral lines. In Fig.~\ref{fig:sp.synph}, we show the modeled spectra 
 at 7d, 31d and 61d including the common set of contributing species (\Hi, \Hei\ etc) and additional 
 set of species \Nii; \Feii; \Tiii; \Scii; \Caii; \Baii; \Nai; \Siii; \Oi\ and  \Ni\ respectively. It can 
 be seen that model spectra is able to match and reproduce most of the 
 features present in the observed spectra. The optimization of $v_{ph}$ is done to match the absorption
 minima of \Feii\ multiplet (4924, 5018, 5169\AA) during plateau phase and of \Hei\ 5876\AA\ line 
 during early phase spectra. 
 

\begin{table*}
  \caption{The best-fit blackbody continuum temperature ($T_{bb}$) and the line velocities 
           of \ha, \hb, \Feii\ (4924\AA, 5018\AA, 5169\AA) and \Hei\ 5876\AA\ as estimated 
           from the \synow\ modeling of the observed spectra of \sn. Velocities derived 
           using lines of \Feii\ or \Hei\ are taken to represent the velocity of photosphere ($v_{\rm phm}$).} 
  \label{tab:synow}

  \begin{tabular}{ccccccc } \hline UT Date & Phase$^{a}$ & $T_{bb}^{b}$ & $v$(\Hei)& $v$(\Feii)& $v$(H$_\alpha$) & $v$(H$_\beta$)\\
     (yyyy-mm-dd)&(day)& (kK) &$10^{3}$ \kms& $10^{3}$\kms& $10^{3}$\kms& $10^{3}$\kms\\ \hline
    2012-03-22 &   7  & 16.5 & 11.2 &   -   & 12.8 & 11.2  \\
    2012-03-23 &   8  & 15.5 & 10.7 &   -   & 12.8 & 11.0  \\
    2012-03-27 &   12 & 14.0 &  9.0 &   -   & 11.7 &  9.7  \\
    2012-03-30 &   15 & 13.5 &  8.65&   -   & 10.6 & 9.3  \\
    2012-03-31 &   16 & 12.5 &  8.5&  8.6 & 10.4 &  9.2  \\
    2012-04-04 &   20 & 10.5 &   -   &  7.7 & 10.0 &  8.6  \\
    2012-04-10 &   26 &  7.5 &   -   &  6.55&  8.8 &  7.3  \\
    2012-04-15 &   31 &  7.5 &   -   &  5.6 &  7.9 &  6.6  \\
    2012-04-29 &   45 &  5.5 &   -   &  4.5 &  6.4 &  5.2  \\
    2012-05-09 &   55 &  5.2 &   -   &  4.15&  5.5 &  4.8  \\
    2012-05-15 &   61 &  5.2 &   -   &  3.5 &  5.0 &  4.2  \\
    2012-05-09 &   66 &  5.2 &   -   &  3.5 &  5.0 &  4.2  \\
    2012-06-27 &  104 &  4.9 &   -   &  2.9 &  3.8 &  3.0  \\
    2012-12-10 &  270 &  3.9 &   -   &  1.6 &  3.4 &  1.6  \\ \hline 
                                    
  \end{tabular}
\begin{flushleft}
  $^{a}$ With reference to the time of explosion JD 24546002.59\\ 
  $^{b}$ Best-fit blackbody termperature of photsphere to match the continuum in observed spectrum.\\ 
\end{flushleft} 
\end{table*}

 The 7d spectrum shows well-matched continuum with P-Cygni profiles of \ha, \hb, \hg, \hd\ and \Hei. 
 The suspected absorption due to high-velocity component of \hb\ and \Hei\ are explored further by
 introducing nitrogen, and the \synow\ modelling appears to be consistent with the absorption dips 
 near 4512\AA, 4834\AA, 5528\AA\ which correspond to \Nii\ lines of rest-wavelength 4623\AA, 5029\AA\ and 5679\AA\ 
 respectively. The \Nii\ lines of 4623 and 5679\AA\ are also seen in 8d 
 spectrum (Fig.~\ref{fig:sp.lit}) and they disappear thereafter. Using full-\nlte\ modelling 
 with \phoneix\ of 7d spectrum of SN 1999em, \citet{1990A&A...237...79B} 
 have reproduced absorption features using \Nii\ lines indicating presence of 
 enhanced nitrogen and helium. We conclude that early-time spectral features of \sn\ were 
 very much similar to normal IIP SNe 1999em as well as 1999gi. We however, note that in 5d spectrum 
 of bright type IIP SN 2007od, the presence of absorption dips near 4500\AA\ \& 4800\AA\ and absence of 
 dip near 5500\AA\, favored presence of HV components \citep{2012MNRAS.422.1178I}. The \Hei\ feature 
 in \sn\ disappears at $\sim$16d and a prominent \Nai\ D feature begin to emerge from 26d at similar 
 location. In plateau-phase spectra, we have also incorporated ions \Feii, \Scii, \Siii, \Caii\ and heavy 
 ions \Tiii, \Baii\  of which most are present as blended or weaker features in the spectra. 

\begin{figure}
 \centering
 \includegraphics[width=4.25cm]{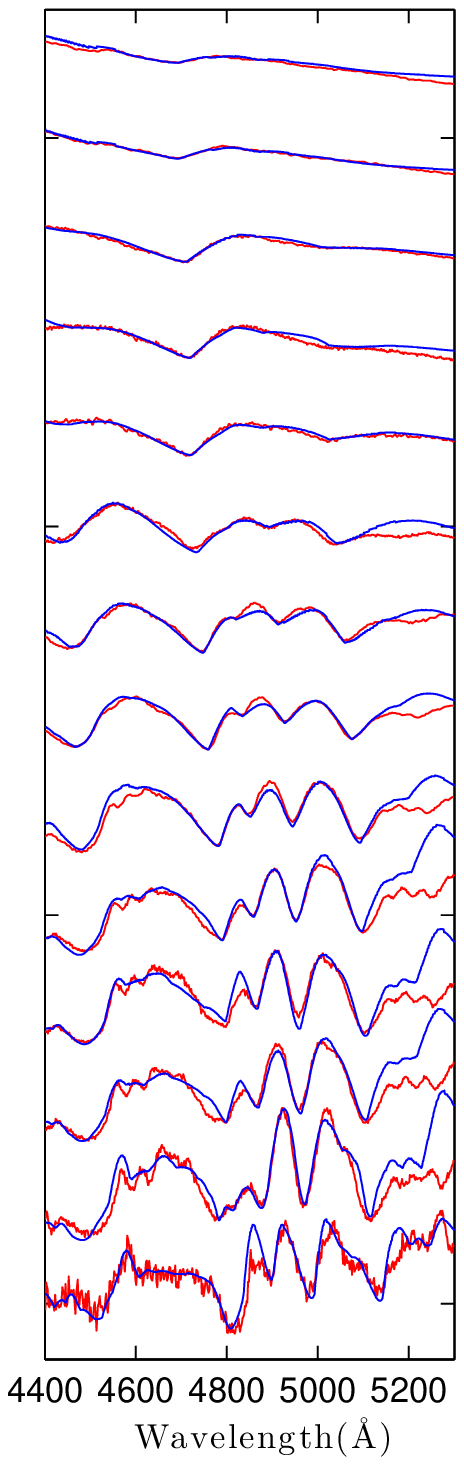}%
 \includegraphics[width=4cm]{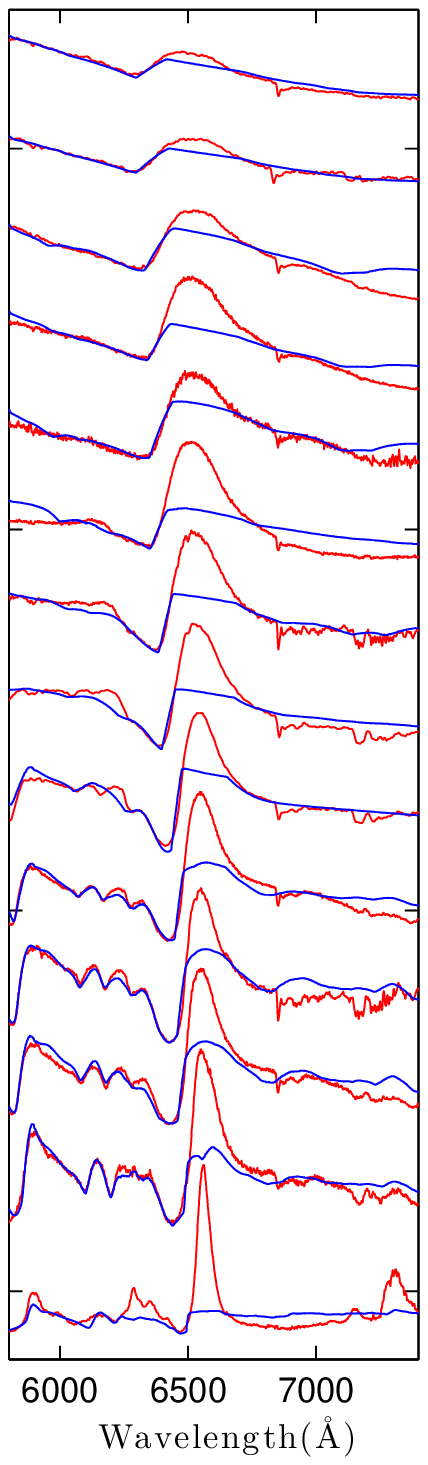}%
 \caption{\synow\ modelling (blue) for \Feii\ multiplet (left) and \ha\ (right) profiles over the observed spectra (red) are shown. The spectral 
          evolution is shown top to bottom at 14 phases from 7d to 270d. Detached \ion{H}\ is used to 
          fit the absorption dip. In addition to  \Feii\ and \Hi; other ions (\Scii, \Baii, \ion{Si}{ii} and \Nai,
          \Tiii\ are also incorporated in model to fit some weaker features, specially at later phases. The phases of plotted spectra are 7d, 8d, 12d, 15d, 16d, 20d, 26d, 31d, 45d, 55d, 61d, 66d, 104d and 270d (top to bottom).}
 \label{fig:sp.synall}
 \end{figure}

 The lines of \Siii\ at 6347\AA\ and 6371\AA\, appearing as a blend at 6355\AA\ can be seen 
 in the bluer wing of broad P-Cygni \ha\ absorption, which starts to appear at about
 26d when $T_{bb}$ drops down to about 7500 K and it can been seen until 
 104d (Fig.~\ref{fig:sp.all}, \ref{fig:sp.synph},\& \ref{fig:sp.synall}). 
 This feature is also seen in the spectra of SN 1999em at similar phases, though it has been 
 identified as a high-velocity component of \ha\ \citep{2002PASP..114...35L}. In \sn\ this feature 
 is fitted well with \Siii\ 6355\AA\ blend in 31d and 61d spectra for the photospheric velocity at 
 respective phases and hence we rule out the possibility of this being a high-velocity component.  
 We note that the \Siii\ feature is also identified in the hotter early time ($<$ 10d) spectra 
 of SN 1999em, which we do not see in \sn.

 The feature marked with C in Fig.~\ref{fig:sp.all} is analyzed further using \synow\ modelling.
 We find that it is most probably a high velocity component of \hb\ blended with \Baii\ 
 and \Tiii\ lines. A careful inspection of model fits in Fig.~\ref{fig:sp.synall}, it can be seen that
 this feature actually starts to appear much earlier than 104d. At 55d, a very faint impression 
 of this feature appears which 
 does not fit with the \synow\ model feature of \hb\ and continues to become increasingly stronger 
 until our last plateau spectra at 104d. At 104d (modeled with \hb\ regular and high velocity 
 component blend) this feature became strong enough to suppress the existing regular \hb\ feature 
 and appear as a entire separate absorption component. By \synow\ modelling the 104d spectra, none 
 of the independent atomic species and single \hb\ component could be accounted for this feature. 
 Only a high velocity \hb\ component of $\sim5200$ km~s$^{-1}$ along with blend of \Baii\ and \Tiii\ 
 lines could match the feature, whereas the regular \hb\ component remains in line with the existing 
 velocity trend at a velocity of $\sim3000$ \kms\ only. Similar problem was also encountered 
 with 104d \ha\ feature which is not much apparent by eye inspection, but single \ha\ component 
 in model spectra was unable to fit the broadened \ha\ absorption dip in the observed spectra. 
 Although no separate absorption dip was seen here, but the absorption feature of \ha\ is somewhat 
 broadened which only a blend of high velocity component along with regular \ha\ can fit.
 This high velocity component is found to be at $\sim5800$ \kms\ whereas 
 the regular \ha\ component is at $\sim3800$ \kms\ which falls in line with the existing \ha\ 
 velocity evolution. It is also to be noted that velocity difference of high velocity component 
 with the regular component is $\sim2000$ \kms\ for both \ha\ and \hb, which might indicate 
 that the origin of both high velocity component is from a single H-layer whose velocity 
 is $\sim2000$ \kms\ higher than the actual layer responsible for photospheric velocity.
 A careful examination of \hb\ profiles in Fig.~\ref{fig:sp.lit} at phase ~60d indicate that
 the feature similar to C is absent in the spectra of SNe 1999em and 1999gi, however, this
 feature is present in spectrum of SN 2004et. The presence of HV components of \ha\ and \hb\ 
 have been observed through out photospheric phase until 105d in a bright ($M_{V}=-17.67$) 
 type IIP SN 2009bw \citep{2012MNRAS.422.1122I} suggestive of interaction 
 between the SN ejecta and the pre-existent circumstellar material (CSM). We, therefore, suggest
 that the broadening of \hb\ and possibly \ha\ during 55d to 104d indicate ejecta-CSM interaction
 in \sn.

 The presence of absorption feature due to \Oi\ line at 7774\AA\ is clearly reproduced by \synow\ 
 in the 61d spectrum. The \Oi\ line begin to appear at 31d and it is clearly seen until 104d.
 The \Ni\ 8130\AA\ line is marginally detected as its identification is affected by the poor 
 SNR. The absorption feature seen $\sim$9000\AA\ in 55d and 66d spectra (see Fig.~\ref{fig:sp.all}) 
 are most likely due to \Ci\ 9061\AA. The appearance of absorption features in the plateau-phase
 spectra due to \Oi\ and occasionally also due to \Ci\ and \Ni\ have also been observed in normal 
 type IIP SN 1999em \citep{2002PASP..114...35L}, bright IIP SN 2009bw \citep{2012MNRAS.422.1122I}, 
 subluminous IIP SNe 2009md \citep{2011MNRAS.417.1417F}, 2005cs \citep{2009MNRAS.394.2266P} and
 2008in \citep{2011ApJ...736...76R}. Several weak absorption features 
 viz. \Feii\ 4629\AA, \Scii\ 4670\AA, \Feii\ 5276\AA\ and \Feii\ 5318\AA\ are identified.


\subsection{Evolution of spectral lines} \label{sec:sp.line}

\begin{figure}
 \centering
 \includegraphics[width=2.33cm]{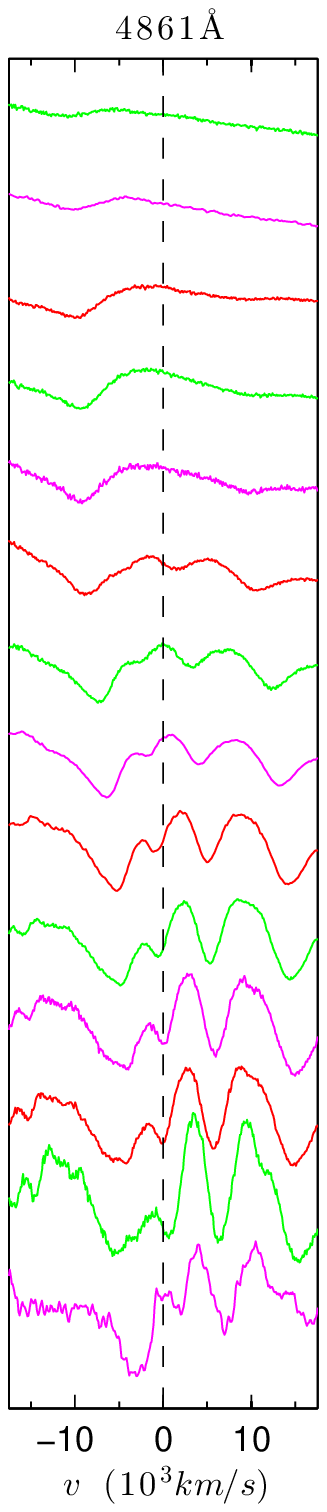} \hskip -4mm%
 \includegraphics[width=2.33cm]{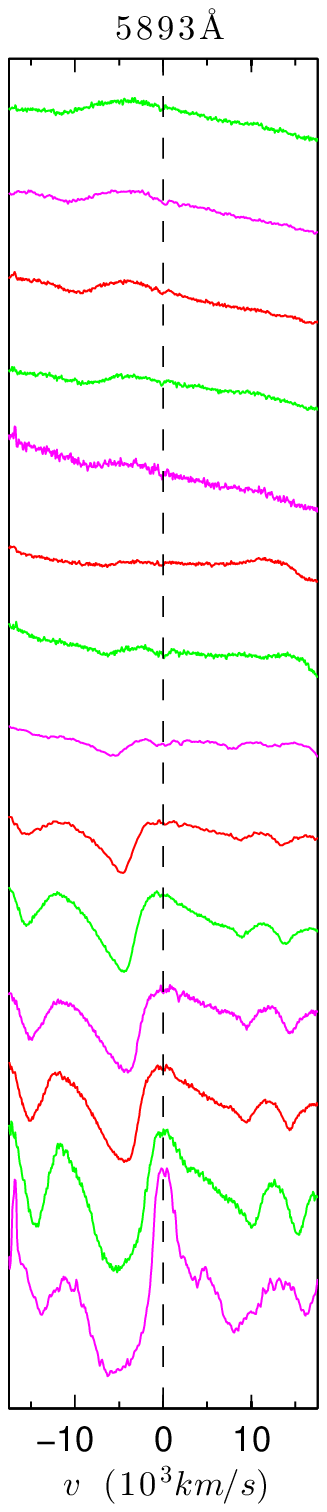} \hskip -4mm%
 \includegraphics[width=2.33cm]{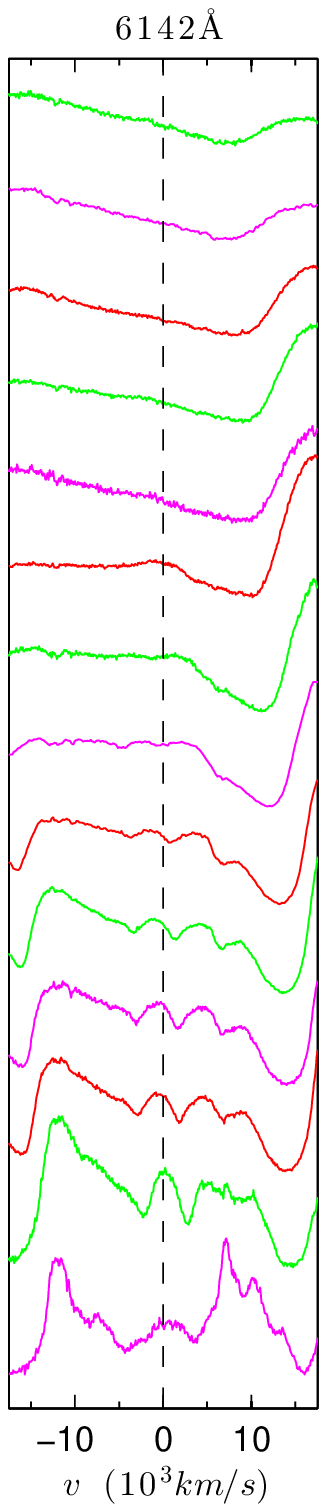} \hskip -4mm%
 \includegraphics[width=2.33cm]{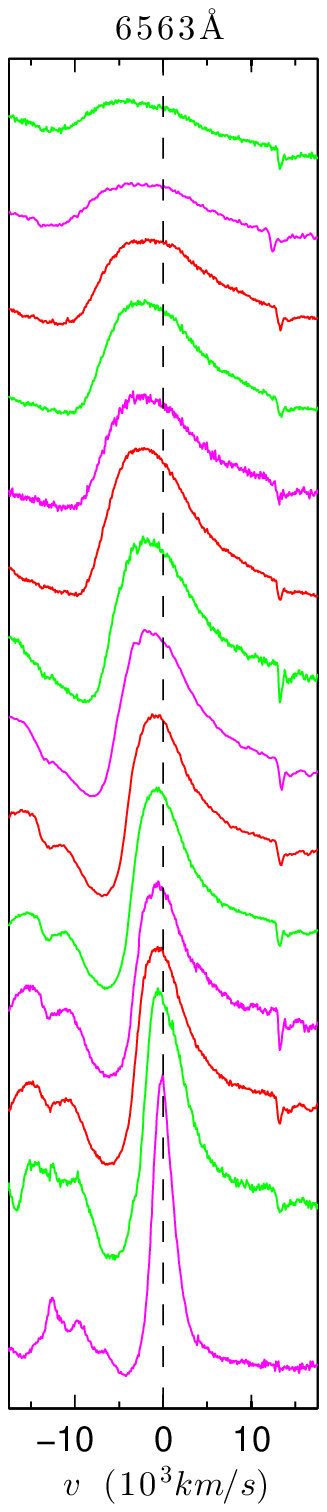}%
 \caption{Evolution of line profiles for \hb, \Nai\ D, \Baii\ and \ha\ are plotted at 14 phases from 7d to 270d top
          to bottom. The zero-velocity line is plotted with dotted line and the corresponding rest 
          wavelength is written on top. The phases of plotted spectra are 7d, 8d, 12d, 15d, 16d, 20d, 26d, 31d, 45d, 55d, 61d, 66d, 104d and 270d (top to bottom).}
 \label{fig:sp.line}
 \end{figure}

 The evolution of spectral features provide important clues about the interaction of expanding ejecta 
 with the circumstellar material, formation of dust in the ejecta and geometrical distribution of 
 ejecta. To illustrate the evolution of individual lines, a portion of spectrum is plotted in 
 Fig.~\ref{fig:sp.line} from 7d to 207d in velocity domain corresponding to rest 
 wavelengths of \hb\ 4861\AA, \Nai\ D 5893\AA, \Baii\ 6142\AA\ and \ha\ 6563\AA. 
 Broad \ha\ P-Cygni (FWHM $\sim$ 17000 \kms) is prominent from as early as first 
 epoch 7d spectra and is present all throughout which evolves into narrow 
 feature (FWHM $\sim$ 2500 \kms at 270d) with time. Apart from blue-shifted absorption 
 trough of the P-Cygni, which is the indicator of expansion velocity, the emission peak 
 is also blue shifted by 4500 \kms\ at 7d (2700\kms at 15d, 2100 \kms\ at 31d), it almost 
 disappears by 66d (600 \kms) and settles to rest in late plateau spectrum of 104d (300 \kms). 
 The blue shift of emission peak in early time (plateau) spectra is also seen 
 in \hb\ and such features are similar with that observed in other type II SNe 
 namely SNe 1999em \citep{2003MNRAS.338..939E}, 2004et \citep{2006MNRAS.372.1315S} and 
 1987A \citep{1987A&A...182L..29H} and it has been 
 theoretically explained as being due to scattering of the photons from the receding part of
 the ejecta \citep{1988SvAL...14..334C,1990sjws.conf..149J}.

 \begin{figure}
 \centering
 \includegraphics[width=8.5cm]{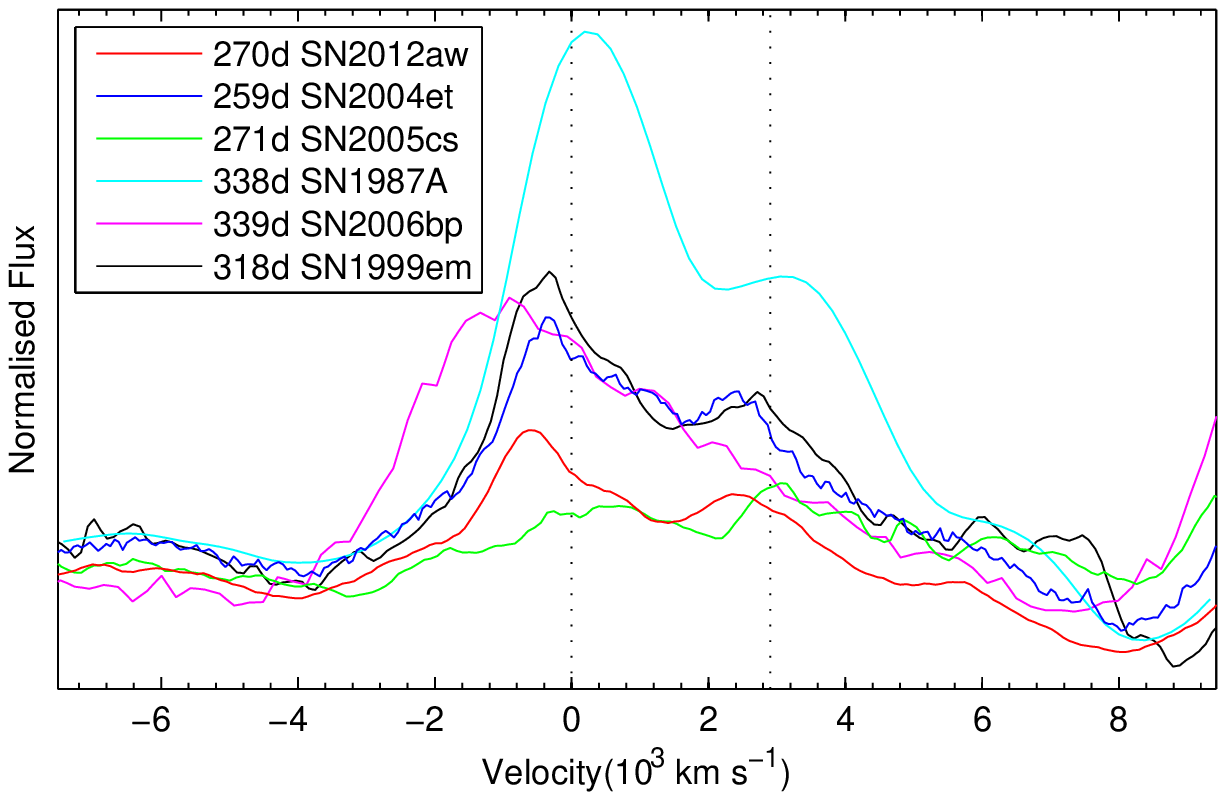}
 \includegraphics[width=8.5cm]{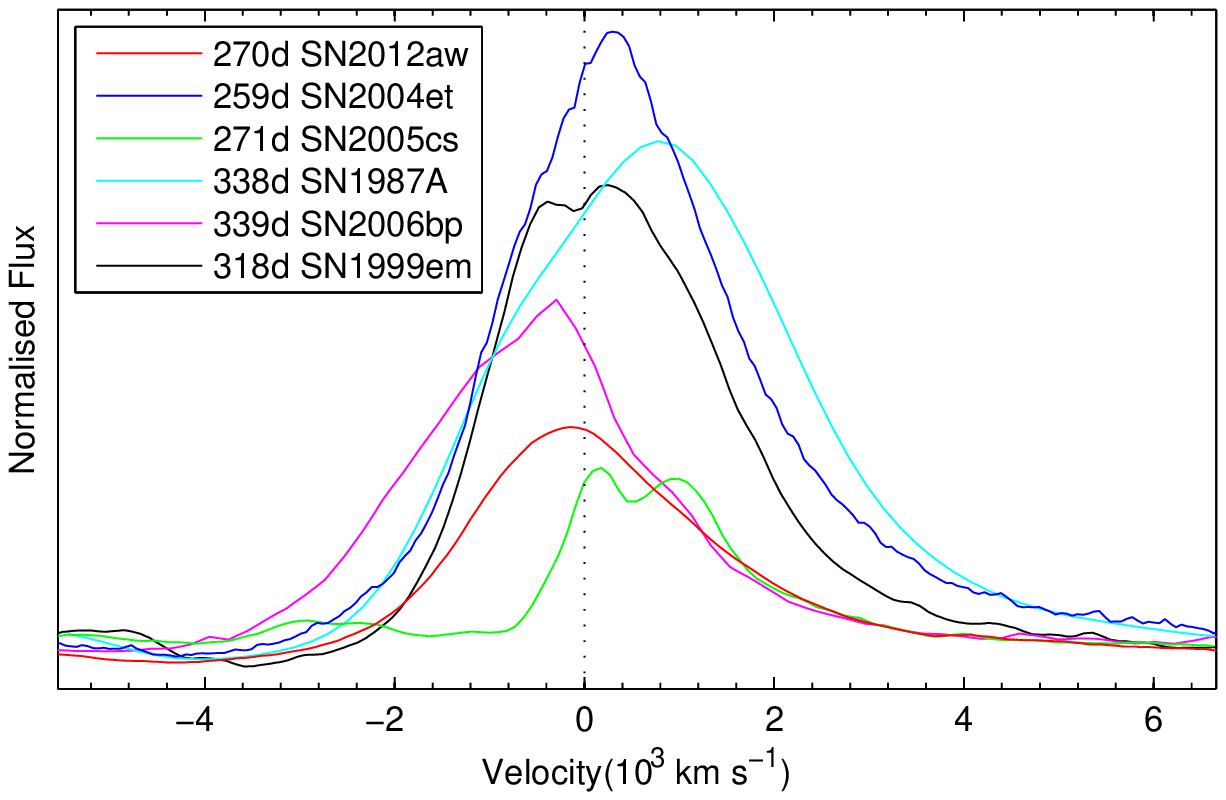}%
 \caption{The velocity profiles of \Oi\ doublet (left) and \ha\ are compared with other SNe from 
          literature. The spectra have been corrected for recession velocities of respective host galaxies. The
          sources for the spectra are : SN 1987A \citep{1995ApJS...99..223P}, SN 1999em \citep{2002PASP..114...35L}, SN 2004et \citep{2006MNRAS.372.1315S}, SN 2005cs \citep{2009MNRAS.394.2266P},
          SN 2006bp\citep{2007ApJ...666.1093Q}.}
 \label{fig:sp.veloh}
 \end{figure}

 A comparison of nebular phase 270d \ha\ line profile with 
 other SNe (Figs.~\ref{fig:sp.lit},~\ref{fig:sp.veloh}) indicate that the \ha\ 
 line flux was lower than that of SNe 1999em and much lower than 2004et. Whereas the FHWM 
 of \ha\ appears to or similar (i.e. $\sim$2600 \kms) for these three SNe, suggesting that 
 the expansion velocity of line-emitting region have been quite similar. Moreover, \ha\ appears 
 to be quite symmetric around zero velocity indicating a spherically symmetric 
 distribution of hydrogen envelope. This is in contrary to that observed for SN 2006bp, whose
 profile is highly asymmetric suggesting that explosion have been quite asymmetric or the 
 dust is forming. The blue shift in \ha\ emission peak can also result from preferential attenuation of
 red wings due to formation of dust in the ejecta. The indication of dust formation is seen in 
 SN 1999em after 500d \citep{2003MNRAS.338..939E} and in SN 2004et 
 after 320d \citep{2006MNRAS.372.1315S,2010MNRAS.404..981M}. The presence of 
 blueshift ($\sim$ 1500 \kms) in \ha\ emission peak have been seen from 226d to 452d spectra 
 of IIP SN 2007od \citep{2012MNRAS.422.1122I} and this has been explained by presence of
 dust in the ejecta. We note that observational signatures related to early dust 
 formation and the ejecta-CSM interaction for \sn\ are unlike that of 
 bright ($M_{V} \sim -18$ mag) IIP SNe 2007od, 2009bw and 2006bp.      

 The \Hei\ 5876\AA\ is prominent from early 7d spectra which disappear rapidly and can be 
 traced only upto 16d. Several metal features start to appear from 26d. The \Siii\ 6355\AA\ feature is 
 found to appear on the blue wing of \ha\, at $\sim -12000$\kms\ and evolved monotonically until 
 last plateau phase spectra at 104d. The \Nai\ D doublet 5890, 5896\AA\ feature start to appear 
 at position very close to earlier \Hei\ feature, and it persists upto nebular phase at 270d. 
 The location of \Nai\ D emission peak at zero velocity indicates an almost spherical 
 distribution of the Na in the ejecta material. The absorption dips due to s-process 
 elements \Baii\ 6142\AA\ and \Scii\ 6248\AA\ start to appear and strengthens until the last 
 plateau spectra 104d. The \Oi\ lines start to appear at 104d and it is seen clearly 
 in 270d spectrum. Similar to \ha\ line, the \Oi\ line flux is too smaller than that of 
 SNe 1999em and 2004et, though the terminal velocity is of similar 
 order (see Fig.~\ref{fig:sp.veloh}).


\subsection{Ejecta velocity} \label{sec:sp.vel}

 The expansion velocity of photosphere ($v_{\rm ph}$) coupled with the mass of ejected matter
 $(M_{\rm ej})$ provides strong constraints on the kinetic energy ($E_{\rm kin}$) of explosion. 
 The photosphere represents the optically thick and mostly ionized part of the ejecta which emits 
 most of the continuum radiation as a ``diluted blackbody". This photosphere is located
 in a thin spherical shell where electron-scattering optical depth of photons
 is $\sim$ $ ^2/_3 $ \citep{2005A&A...437..667D}. In type IIP SNe, no single measurable 
 spectral feature is directly connected with the true velocity of photosphere, 
 however, during the plateau-phase, it is best represented by blue-shifted absorption components
 of P-Cygni profiles of \Feii\ at 4924\AA, 5018\AA\ and 5169\AA, while in early-phase
 lines of \Hei\ 5876\AA\ or \hb\ act as a good proxy (see TV12 for a detailed 
 review on estimating photospheric velocities of type IIP SNe). We can 
 estimate velocities either by measuring Doppler-shift of 
 the absorption minima using \splot\ task of \iraf\ or by modelling the observed spectra 
 with \synow. The later gives better estimate of $v_{\rm ph}$ as it can take care  and reproduce line blending, 
 however, for the sake of comparing velocities of \sn\ with other SNe in literature, we use both the 
 methods in this work and the corresponding velocities are denoted as $v_{\rm pha}$ and $v_{\rm phm}$
 respectively. Apart from \Feii\ lines, the expansion velocities of line-forming layers for \ha, \hb, \Hei, and \Scii\ are also 
 estimated.

 Using \synow\ we obtain the best fit locally, for employing the whole wavelength range 
 may lead to over- or under-estimate of velocities due to formation of lines
 at different layers. Fig.~\ref{fig:sp.synall} shows best fitted profiles over $\sim$ 1000\AA\ wide
 wavelength regions around \ha\ and \Feii\ along with \hb\ features. It can be seen that the 
 best-fit model spectra are able to reproduce the absorption components 
 of \hb\,, \Feii\ 4924\AA, \Feii\ 5018\AA, and \Feii\ 5169\AA\ simultaneously. 
 At early phases from 7d to 16d, the best-fit model velocity for \Hei\ 5876\AA\ line
 is also estimated. The model-derived velocities are listed in Table~\ref{tab:synow} and the 
 value of $v_{\rm phm}$ represent \Hei\ line until 15d and \Feii\ lines at phases thereafter. 
 The typical uncertainty in velocities estimated by deviation seen visually from best-fit
 absorption troughs by varying $v_{\rm phm}$ is $\sim$~150 \kms. This is consistent with the values 
 obtained using automated computational techniques viz. $\chi^{2}$-minimization and cross-correlation 
 methods (TV12).

\begin{figure}
 \includegraphics[width=8.5cm]{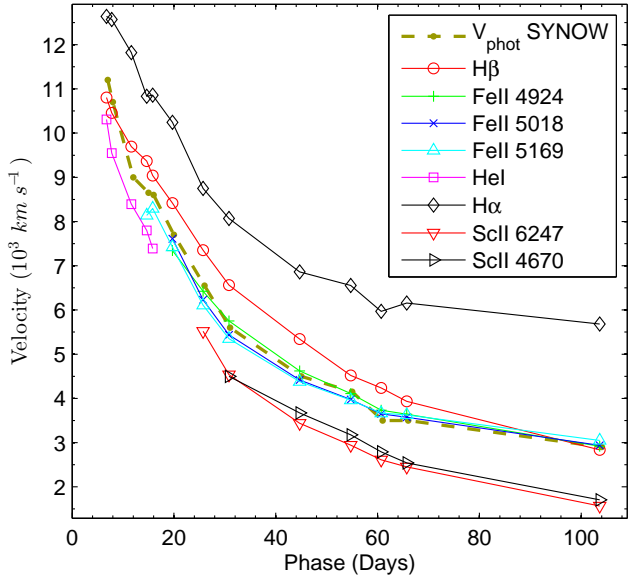}%
 \caption{Line velocity evolution of \ha, \hb, \Hei, \Scii\ and \Feii. The velocities are estimated using 
          Doppler-shift of the absorption minima. The expansion velocity of photosphere ($v_{\rm phm}$) estimated from
          \synow\ fits of \Hei\ line until 15d and simultaneous fits for \Feii\ lines at later phases
          (see Table~\ref{tab:synow}) are overplotted for comparison.}
 \label{fig:sp.velall}
 \end{figure}

 The expansion line velocities of \ha, \hb, \Hei, \Feii\ (4924\AA, 5018\AA\ and 5169\AA) components,
 \Scii\ 6246\AA\ and \Scii\ 4670\AA\ blend 
 has also been determined using \iraf\ by fitting the absorption trough with a Gaussian
 function and these are shown in Fig.~\ref{fig:sp.velall}. It can be seen that \Hi\ (\ha, \hb) lines
 formed at larger radii (i.e. higher optical depths) than \Hei, while the \Feii\ lines are formed at 
 lower radii. The \Scii\ lines are formed at even lower optical depths and the velocities 
 derived using \Scii\ 4670 and 6247\AA\ at phases 26d onwards are systematically lower by $ \sim $1000 \kms\ from 
 that of \Feii\ lines or of $v_{\rm phm}$. During 16d to 104d, it can be seen that 
 velocities derived using individual lines of \Feii\ are same within errors and these values are 
 further in agreement with that determined using \synow\ (i.e. simultaneous fits 
 to \Feii\ lines). At early phases ($<$ 15d), the \synow\--determined values from \Hei\ are found to
 be consistently higher by $\sim$ 1000\kms\ from that determined using absorption minima, however, at
 phases 7d and 8d these values are consistent with that obtained from \hb\ absorption minima.
 The \ha\ velocity is higher than $v_{\rm phm}$ by $\sim 2000$\kms at early phases and 
 by $\sim 1000$ \kms at later phases. As noted by TV12, we confirm
 that the line velocities of \hb\ are in agreement with that of $v_{\rm phm}$ above
 10000\kms\,, while below this it is consistently higher than that determined using \Feii\ lines.
 The value of $v_{\rm phm}$ falls from 2900 \kms\ at 104d to 1600\kms\ at 270d.

\begin{figure}
\centering
\includegraphics[width=8.5cm]{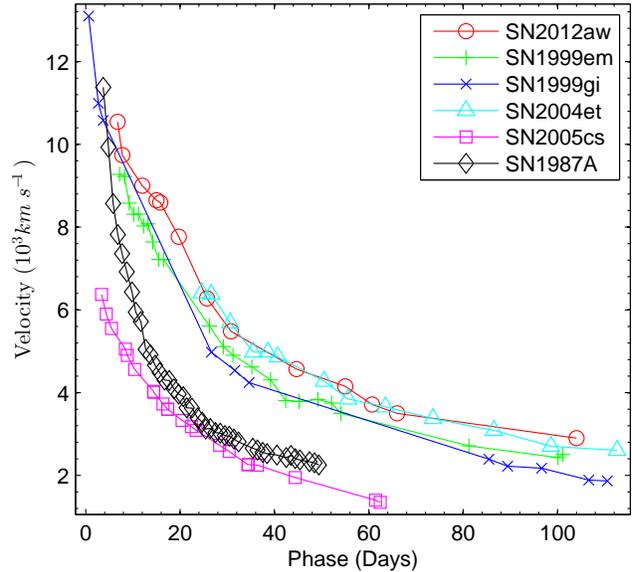}%
\caption{The evolution of photospheric velocity ($v_{\rm ph}$) of {\sn} is compared with other well-studied supernovae.
         The ($v_{\rm ph}$) plotted here are the absorption trough velocities (average of \Feii\ lines at late 
         phases and \Hei\ at early phases.}
\label{fig:sp.velph}
\end{figure}

 Fig.~\ref{fig:sp.velph} shows the comparison of photospheric velocity of \sn\ with other well-studied 
 SNe 1999em, 1999gi, 2004et, 1987A and 2005cs. For this comparison purpose the absorption trough 
 velocities (average of \Feii\ lines at late phases and \Hei\ at early phases) have been used as 
 the \synow\--derived velocities are not available for all the SNe  
 considered for comparison\footnote{Barring SN 1999em, taken from \citet{2002PASP..114...35L}, 
 the velocities for all other comparison SNe are determined in this work using spectra available at 
 \suspect\ {http://suspect.nhn.ou.edu .}}. The velocity evolution of \sn\ is similar to the 
 normal type IIP SNe 2004et, 1999em and 1999gi; thought it is strikingly different than that 
 of subluminous SN 2005cs and the type II-peculiar SN 1987A. The velocities of SN 2005cs are extremely 
 less than {\sn} at all phases by $\sim3000$ \kms, whereas the profile of SN 1987A shows sharp 
 decline in early phases ($<$15 days) and comparatively slower decline in later phases. The entire 
 photospheric velocity profile of {\sn} is identical to 2004et at all phases whereas it is consistently 
 higher than that of SNe 1999em and 1999gi by $\sim600$ \kms\ at all phases. We note that while comparing \Scii\ 
 6247\AA\ absorption velocities of SNe 2004et and 1999em, no difference is seen in velocity 
 evolution \citep{2010MNRAS.404..981M}, however, velocities obtained using \synow\ model fits to 
 the \Feii\ lines result in systematical higher velocities for SN 2004et (TV12). Similarly a comparison 
 of expansion velocities of \ha\, and \hb\ with other SNe indicate that these too are systematically 
 higher than that observed for SNe 1999em and 1999gi; and are comparable with that observed for SN 2004et.
 For example, at phases 11d, 30d, and 50d respectively, the velocities of \hb\ are 11000, 7400, 
 and 5800 \kms\ for SN 2004et (TV12); 10000, 6600, and 5000 \kms\ 
 for SN 2012aw (Table~\ref{tab:synow}); 8400, 5200 and 3500 \kms\ for SN 1999em (TV12).

\section{Characteristics of explosion} \label{sec:char}


\subsection{Explosion energy} \label{sec:char.energy}

 The radiation-hydrodynamics simulations provided by \cite{2010MNRAS.408..827D} for core-collapse 
 SNe generated artificially by driving a piston at the base of the envelope of a rotating and 
 non-rotating red-supergiant progenitor stars, suggest that the $v_{\rm ph}$ at 15d after shock 
 breakout is a good and simple indicator of the explosion energy ($E_{0}$), no matter 
 what the initial mass is. 
 For non-rotating solar metallicity models
 with progenitor masses 11-30\msun, a simulated plot of $v_{\rm ph, 15d}$ and the velocity at the 
 outer edge of the oxygen-rich shell $v_{\rm ej, O}$ (see Fig.4 in their paper) for different 
 energy of explosion ranging from 0.1 to 3 foe (1 foe =$10^{51}$ erg), indicate that the  
 for \sn\ corresponding to the value of $v_{\rm ph, 15d}$ = 8650 \kms (see Table~\ref{tab:synow}), 
 the value of $E_{0}$ lies in the range 1-2 foe. The value of $v_{\rm ph, 15d}$ for SN 1999em 
 is 7650 \kms\ whereas for SN 2004et it is 8800 \kms\ (TV12) and hence assuming similar nature
 of progenitor star, the strength of explosion for \sn\ should have been similar 
 to SN 2004et but higher than that of SN 1999em. We note that a detailed radiation-hydrodynamic
 simulations of bolometric light-curves obtain value of $E_{0}$ of $1.3\pm0.3$ foe for 
 SN 1999em \citep{2007A&A...461..233U} and 
 of $2.3\pm0.3$ foe for SN 2004et \citep{2009A&A...506..829U}. 

 Furthermore, the simulated models by \cite{2010MNRAS.408..827D} when combined with the width
 of observed nebular-phase \Oi\ 6300-6364\AA\ line, can also be used to place an upper limit of the 
 progenitor main-sequence mass. In nebular-phase spectra, an assessment of $v_{\rm ej, O}$ can be 
 made by measuring half-width half-maxima (HWHM) of \Oi\ feature and for \sn\ it is found
 to be $\sim$ 1340 \kms\ at 270d and for comparison SNe
 2004et, 1999em, 1987A, it is $\sim$1300\kms\, $\sim$1200\kms\ and $\sim$1400\kms\ 
 respectively (see Fig.~\ref{fig:sp.veloh}). Hence the simulations 
 for non-rotating models suggests an upper mass of 15\msun\ for the progenitor star of SN 2012aw. Rotating models 
 suggest lower-mass star. This upper mass limit 
 for \sn\ is consistent with that determined by \cite{2012ApJ...759...20K} using data on pre-supernova 
 stars and the stellar evolutionary models. We however note that the simulations 
 of \citet{2010MNRAS.408..827D} assume no mixing of ejecta whereas the study of velocity distribution 
 in line profile shapes of nebular lines by \citet{2012MNRAS.420.3451M} indicate that mixing in ejecta 
 is likely to have occurred in type IIP SNe. Consequently the width of \Oia\ lines may not represent 
 true velocity of the oxygen-rich zones and hence the simulations-derived progenitor mass may be regarded
 as first hand estimates.


\subsection{Mass of progenitor star} \label{sec:char.mass}

 The mass of progenitor star can be estimated using nebular-phase emission line
 of \Oi\ 6300, 6364\AA\ doublet as a detailed nucleosynthesis yields of stellar 
 evolution/explosion models indicate 
 that the core mass of metals in the inner ejecta is found to scale with 
 the zero-age main-sequence mass of the progenitor \citep{1995ApJS..101..181W}. 
 Additionally, we also consider that in type II SNe, the nucleosynthesis yield is 
 largely unaffected by late-time evolution of supernova ejecta. The mass of oxygen in type II SNe can be 
 estimated by analysis of emission from the nebular phase \Oia\ doublet which is mainly 
 powered by $\gamma$-ray depositions and more than half of the \Oia\ doublet luminosity 
 during nebular phase is contributed by newly synthesized oxygen in the 
 ejecta \citep{2012A&A...546A..28J}. 
 Also, considering the fact that the SNe 2004et and 1987A have similar ejecta velocities 
 and \nickel\ masses; and the mass of oxygen for SNe 2004et and 1987A is modeled quite accurately
 and is available for comparison.
 We however note that the comparison with SN 2004et is more relevant as it is spectroscopically and photometrically similar to SN 2012aw.
  Following \citet{2003MNRAS.338..939E}, the luminosity 
 of \Oia\ doublet can be written as:
 \begin{eqnarray*} 
  L_{6300} = \eta \frac{M_{\rm O}}{M_{\rm ex}} L(\cobalt)
 \end{eqnarray*}
 where  $M_{\rm O}$ is the mass of oxygen, $M_{\rm ex}$ is the `excited' oxygen mass in which
 bulk of decay energy is deposited, $\eta$ is the efficiency of transformation of decay 
 energy deposited into \Oia\ doublet radiation and $L(\cobalt)$ is the luminosity of
 \cobalt\ which is directly proportional to the mass of ejected \nickel. 

 The \Oi\ luminosity of $\sim 1.36 \times 10^{39}$ \ergs\ is obtained for \sn\ by 
 integrating the flux within local minima (6214 -- 6471\AA) and subtracting the local continuum in that region. 
 Similarly a value of $\sim 1.64 \times 10^{39}$ \ergs\ is estimated for SNe 2004et at 270d by 
 interpolating the value using 259d and 301d spectra taken from \citet{2006MNRAS.372.1315S}.
 Assuming that in both SNe 2004et and 2012aw, $\eta$ and $M_{\rm ex}$ are similar and given 0.83 times lower 
 luminosity of \Oia\ doublet in \sn\ at 270d
 and equal \nickel\ mass (\S\ref{sec:lc.nick}), we derive a rough estimate of oxygen in \sn\ to be a 
 factor 0.83 lower than in SN 2004et. Considering the oxygen mass of SN 2004et as 0.8 \msun, estimated 
 using spectral modelling of nebular phase 
 (140d to 700d) spectra from ultraviolet to mid-infrared \citep{2012A&A...546A..28J},
 the mass for \sn\ translates to be 0.66 \msun\ and using nucleosynthesis yield computations in 
 massive stars (11-40 \msun) by \citet{1995ApJS..101..181W} corresponds to the main-sequence 
 solar metallicity stellar mass of  14-15  \msun.   

 The comparison with \Oi\ luminosity of SN 1987A (a value of $\sim 2.1 \times 10^{39}$ \ergs\ is 
 estimated by 
 interpolating values obtained at 197d and 338d using spectrum taken from \cite{1995ApJS...99..223P}) 
 with SN 2012aw indicate that mass of oxygen in \sn\ to be a factor 0.84 lower than in SN 2004et. 
 Considering the oxygen mass determinations of SN 1987A in 
 range 1.2-1.5 \msun\ \citep{1992ApJ...387..309L,1994ApJ...428L..17C,1998ApJ...497..431K},
 the mass range for \sn\ translates to be 1.0-1.26 \msun\ and using nucleosynthesis 
 computations this corresponds to the main-sequence solar metallicity stellar mass 
 of 17-19 \msun\ \citep{1995ApJS..101..181W}.

 The minimum mass of oxygen can also be obtained independently using the 
 equation \citep{1986ApJ...310L..35U}:
 \begin{eqnarray*}
 \displaystyle{M_{\Oi} =  10^{8}  {F_{\Oi}} {D^{2}} {e^{2.28/T_{4}}} }
 \end{eqnarray*}
 where F$_{\Oi}$ is the \Oi\ doublet flux in units of \ergs, D is the distance to the SN 
 in units of Mpc and T$_{4}$ is the temperature in units of 10$^4$ K. From \cite{1995ApJ...454..472L}, the O 
 temperature of SN 1987A at 300 days was $\sim$ 4200 K. Assuming a similar O temperature for 
 SN 2012aw at a comparable epoch, the oxygen mass for SN 2012aw was calculated for temperatures in 
 the range 3500--4500 K as 0.18 -- 0.77 M$_\odot$ with the corresponding minimum main-sequence masses
 in the range 11-16\msun.



\subsection {Explosion parameters} \label{sec:char.par}


\begin{table*}
  \caption{Explosion parameters of well-studied type IIP SNe. The references for the adopted time
           of explosion ($t_{\rm 0}$), distance ($D$) and reddening $E(B-V)$ are given in 
           Fig.~\ref{fig:lc.abs}. See \S\ref{sec:char.par} for further details.} 
  \label{tab:expar}

  \begin{tabular}{lcccccccc} 
     \hline
 Name& $t_{\rm 0}$& $\Delta t_{\rm p}$& $t_{\rm p}$ & $M^{\rm p}_{V}$& $v_{\rm p}$&            $E_{\rm 0}$& $M_{\rm ej}$& $R_{\rm 0}$ \\
     &  (2450000+)&             (day) &       (day) &        (mag)   &      (\kms)& ($\times 10^{50}$ erg)&      (\msun)&  (\rsun)\\
     \hline
     SN 1999em& 1475.6&   $ 92\pm8$&  $55\pm4$&  $-16.69\pm0.01$&   $3512\pm122$&   $7\pm2$&   $11\pm3$&   $399\pm54$\\
     SN 1999gi& 1522.3&   $ 97\pm8$&  $58\pm4$&  $-16.26\pm0.02$&   $2746\pm217$&   $4\pm1$&   $10\pm3$&   $421\pm99$\\
     SN 2004et& 3270.5&   $ 87\pm8$&  $63\pm4$&  $-17.01\pm0.03$&   $3630\pm142$&   $6\pm2$&   $ 9\pm2$&   $591\pm90$\\
     SN 2012aw& 6002.6&   $96\pm11$&  $57\pm6$&  $-16.67\pm0.03$&   $3631\pm200$&   $9\pm3$&   $14\pm5$&   $337\pm67$\\
     \hline
  \end{tabular}
\end{table*}

%
%

 Accurate estimates of explosion parameters of type IIP SNe require detailed hydrodynamical 
 modelling of their bolometric light-curve \citep[e.g.][references therein]{2011ApJ...729...61B}
 which is beyond the scope of this paper, however the analytical relations connecting the 
 observed parameters (viz. the duration of plateau $\Delta t_{\rm p}$, the mid-plateau absolute 
 magnitude $M^{\rm p}_{V}$ magnitude and mid-plateau photospheric velocity $v_{p}$) with  
 physical parameters of the explosion (viz. the energy of the explosion $E_{0}$, the radius of 
 progenitor star $R_{0}$ and the mass of the ejected matter $M_{\rm ej}$) do 
 exist \citep{1980ApJ...237..541A}. \citet[][LN85]{1983Ap&SS..89...89L,1985SvAL...11..145L} made numerical 
 calibration of these relations for a wide range of observables using a grid of hydrodynamical 
 models for different values of $E_{\rm 0}$, $R_{\rm 0}$ and $M_{\rm ej}$. The applicability of such a  
 relation have been questioned as the ejected masses derived using LN85 relations for large 
 set of IIP SNe \citep[ 14-56  \msun by][]{2003ApJ...582..905H}; 
 \citep[ 10-30  \msun\ by][]{2003MNRAS.346...97N}; are consistently higher than that obtained using direct 
 pre-SN imaging \citep[6-15 \msun by][]{2009MNRAS.395.1409S}. Some of the problems lie in the lack
 of good quality data, simplified physical assumptions and non-inclusion of nickel heating 
 effects \citep[e.g. see][]{2011ApJ...729...61B}, however, these relations are still useful 
 in comparing the relative explosion properties of IIP SNe. 

 For the sake of estimating observed parameters for \sn\ and comparing with other SNe, having similar
 light-curve/spectra behavior, in a consistent manner, we consider nearby  
 normal type IIP SNe 1999em, 1999gi and 2004et for which the distances, reddening
 and time of explosions are known quite accurately and all of these have good photometric 
 and spectroscopic observations during plateau phase. The value of $\Delta t_{\rm p}$  to be
 determined using $M_{V}$ light-curve (see Fig.~\ref{fig:lc.abs}) in a manner described 
 by \cite{2003MNRAS.346...97N} is non trivial, however, a plateau duration of $\sim$ 100 days
 is clearly apparent for all the four SNe. As noted in \S\ref{sec:lc.app}, the V-band light-curve
 of sample SNe peaks between 10-16d post explosion and show a slow linear decline until $\sim$ 110d,
 we have therefore fitted a straight line to the linear part of the plateau to estimate the
 phases at which the slope changes significantly by including two consecutive points at the ends. 
 The value of $\Delta t_{\rm p}$ determined in this way and the phase $t_{\rm p}$ corresponding 
 to mid-point are given in Table~\ref{tab:expar}. Furthermore, at $t_{\rm p}$, we obtain both, the value 
 of $M^{\rm p}_{V}$ by linear interpolation (data from Fig.~\ref{fig:lc.abs}) and the value of $v_{\rm p}$ 
 by third order polynomial interpolation of $v_{\rm pha}$ values shown in Fig.~\ref{fig:sp.velph}, 
 which were determined by absorption minima to \Feii\ lines. The explosion parameters determined
 in this way is listed in Table~\ref{tab:expar} and it can be seen that the properties of 
 \sn\ is very much similar to SNe 1999gi, 1999em and 2004et. 

 The value of $E_{0}$ is close to the 
 standard energy of SNe explosion ($\sim 10^{51}$ erg) and the pre-supernova radius
 is consistent with that of Galactic red supergiant stars measured observationally 
 by \citet{2005ApJ...628..973L}. The value of $M_{\rm ej}$ lies in the range 9-14 \msun\ and 
 accounting for a total of  2\msun\ including mass of neutron star (a possible endpoint of IIP SNe) and 
 the mass loss during 
 red supergiant phase, this corresponds to a main sequence mass of  11-16 \msun.
 It is noted that using \Scii\ 4670\AA\ line for the photospheric velocities in our 
 computation provides lower values of ejected masses. These values are consistent with 
 the values determined using direct imaging of pre-supernova stars \citep{2009MNRAS.395.1409S}.
 Furthermore, it is seen that all these SNe have similar properties in terms of explosion energy, 
 ejected mass and the radius of pre-supernova star and considering uncertainty in estimating
 $\Delta t_{\rm p}$ and $v_{\rm p}$, it is difficult to find any trend in their relative explosion
 parameters, and we need a larger sample of nearby IIP SNe having good-quality data and a uniform approach
 to verify the applicability of analytical relations. For example, we note that using the LN85 
 relations, \citet{2010MNRAS.404..981M} derived the value of $M_{\rm ej}$ for SNe 1999em, 1999gi 
 and 2004et in the range 14-21 \msun\,  and they employed plateau duration in the range 110-120 days 
 and the photospheric velocity derived from  \Scii\ 6246\AA\ line.

\section {Conclusions} \label{sec:sum}
 We present new $UBVRI$ photometric and low resolution spectroscopic observations of a 
 supernova event SN 2012aw which occurred in the outskirt of a nearby ($9.9\pm0.1$ Mpc) 
 galaxy M95. The time of explosion is constrained with an accuracy of a day and the 
 position of SN in the galaxy is consistent with being located in a solar metallicity 
 region. The photometric observations are presented at 45 phases during 4d to 269d while
 the low-resolution (6-12\AA) spectroscopic observations are presented at 14 phases 
 during 7d to 270d. Employing the high-resolution spectrum of \Nai\ D region and the early
 time photometric spectral energy distribution, the value of $E(B-V)$ is constrained
 quite accurately to be $0.07\pm0.01$ mag.  

 The light-curve characteristics of apparent magnitudes, colors and the bolometric
 luminosity is found to have striking similarity with the archetypal IIP SNe 1999em,
 1999gi and 2004et; all showing plateau duration of about 100 days. For all these
 SNe, the light-curve in $V$-band rises to a peak between 10-16d post explosion and then
 follows slow decline during plateau-phase. However, for \sn\, our early time observations 
 clearly detect minima in the light-curve of $V$, $R$ and $I$ bands near 32 days after 
 explosion and this we suggest to be an observational evidence seen for the first time 
 in any type IIP SNe, for the emergence of flux due to onset of recombination phase.
 The value of mid-plateau $M_{V}$ is $-16.67\pm0.04$ for \sn\ lies
 in between the bright IIP SNe ($\sim -18$ mag; e.g 2007od, 2009bw) and the subluminous
 IIP SNe ($\sim -15$ mag; e.g. 2005cs, 1997D). Employing nebular-phase bolometric 
 luminosity, we estimate mass of \nickel\ to be
 $0.06\pm0.01$, similar to the SNe 1999em, 2004et and 1987A.

 The presence and evolution of prominent optical spectral features show striking similarity 
 with the IIP SNe 1999em, 1999gi and 2004et. We have identified and studied the evolution 
 of spectral features using \synow\ modelling. Similar to SNe 1999em, and 1999gi, two peculiar 
 high-velocity components associated with the regular \hb\ and \Hei\ P-Cygni features are 
 seen in the early (7 \& 8d) spectra, indicating early interaction of ejecta with the 
 circumstellar material. However, these absorption features are consistent with 
 being reproduced by invoking \Nii\ lines in the \synow\ modelling. During 55-104d, 
 the absorption profiles of \hb\ and \ha\ is broadened and it only fit by invoking high-velocity
 components showing signs of ejecta interaction with CSM during late plateau phase.
 We note that interaction scenario is consistent with the detection of \sn\ in X-rays
 (0.1-10 keV) during 4-6d \citep{2012ATel.3995....1I} and at 21 GHz radio observations 
 during 8-14d \citep{2012ATel.4012....1S,2012ATel.4010....1Y}      

 The velocity of \ha\ and \Oia\ doublet line-emitting regions in the nebular phase spectrum 
 at 270d is found to be similar to that observed for SNe 1999em and 2004et; and the line profile
 shapes are consistent with being originated from spherically symmetric regions, showing no signs
 of dust formation.  

 The value and evolution of photospheric velocity as derived using \Feii\ lines is found to be 
 similar to SN 2004et, but about $\sim$600 \kms\, higher than that of SNe 1999em and 1999gi 
 at similar epochs. This trend was more apparent in the line velocities of \ha\ and \hb. 
 The comparison of photospheric velocity at 15d with that derived using radiation-hydrodynamics
 simulations of IIP SNe by \citet{2010MNRAS.408..827D} indicate that the energy of explosion is about 
 $1-2\times10^{51}$ erg and this coupled with the velocity of \Oia\ line suggests an upper mass limit 
 of 15\msun\ for a non-rotating solar metallicity progenitor star. We further constrain, the 
 progenitor mass by comparing \Oia\ emission luminosity with the SNe 2004et and 1987A and 
 we find that the core of oxygen mass was smaller than that of SNe 2004et and 1987A; and assuming 
 similar physical conditions, we derive mass of progenitor star about 14-15 \msun.

 We have also estimated explosion parameters using analytical relations of LN85 for SNe
 1999gi, 1999em, 2004et and 2012aw; all having good coverage of photometric and spectroscopic 
 data during plateau phase, in a consistent manner. We find no trend in relative parameters but
 ensemble parameters are found to be consistent with that expected for a normal luminosity 
 IIP SNe, i.e., the explosion energy is consistent with $1\times10^{51}$ erg, the pre-supernova 
 radius is similar to what is expected from a red-supergiant star and the mass of progenitor 
 lies between 11-16\msun. 

 \sn\, along with type IIP SNe 1999gi, 1999em and 2004et forms a golden sample to test
 results from radiation hydrodynamical simulations.

\section*{Acknowledgments}
 We thank all the observers at Aryabhatta Research Institute of Observational
 Sciences (ARIES) who provided their valuable time and support for the
 observations of this event. We are thankful to the observing
 staffs and technical assistants of ARIES 1.3-m Devasthal telescope and we also express our thanks to 2-m IGO, 2-m HCT telescope staffs for their kind cooperation 
 in observation of \sn. We gratefully acknowledge the services of the
 NASA ADS and NED databases and also the online supernova spectrum archive (SUSPECT) which are used to access data and references in this paper. We thank K. Maguire for providing late spectroscopic data of SN 2004et. The authors would also like to thank the anonymous referee for the comments and suggestions which helped in improvement of this manuscript.


\bibliographystyle{mnras}
\bibliography{ms}

\label{lastpage}

\end{document}